\documentclass[fleqn,usenatbib]{mnras}

\usepackage{newtxtext,newtxmath}

\usepackage[T1]{fontenc}
\usepackage{ae,aecompl}
\usepackage{afterpage}
\usepackage{adjustbox}
\usepackage{media9}
\usepackage[utf8]{inputenc}
\usepackage[T1]{fontenc}
\usepackage{parskip}

\usepackage{graphicx}	
\usepackage{amsmath}	
\usepackage{subfig}
\usepackage{comment}
\usepackage{enumerate}

\newcommand{\UK}[1]{\textcolor{red}{\bf comment: #1}}
\newcommand{\hMpc}{{\ifmmode{~h^{-1}{\rm Mpc}}\else{$h^{-1}$Mpc}\fi}}
\newcommand{\Msun}{{\ifmmode{{\rm {M_{\odot}}}}\else{${\rm{M_{\odot}}}$}\fi}}
\newcommand{\gadgetx}{\textsc{Gadget-X }}
\newcommand{\threehundred}{\textsc{The ThreeHundred }}
\newcommand{\disperse}{\textsc{DisPerSE}}
\newcommand{\Amiga}{\textsc{AMIGA}}

\title[Mapping and characterisation of cosmic filaments]{Mapping and characterisation of cosmic filaments in galaxy cluster outskirts: strategies and forecasts for observations from simulations 
}

\author[U. Kuchner et al.]{\parbox{\textwidth}{
Ulrike Kuchner,$^{1}$\thanks{E-mail: ulrike.kuchner@nottingham.ac.uk (uk)}
Alfonso Arag\'{o}n-Salamanca,$^{1}$
Frazer R. Pearce,$^{1}$
Meghan E. Gray,$^{1}$
Agust\'{i}n Rost,$^{1,2}$
Chunliang Mu,$^{1,3}$
Charlotte Welker,$^{4}$
Weiguang Cui, $^{5}$
Roan Haggar,$^{1}$
Clotilde Laigle,$^{6}$
Alexander Knebe$^{7,8,9}$
Katarina Kraljic,$^{5}$
Florian Sarron,$^{1}$
Gustavo Yepes$^{7,8}$
}
\vspace{0.4cm}
\\
\parbox{\textwidth}{
$^1$School of Physics \& Astronomy, University of Nottingham, Nottingham NG7 2RD, UK\\
$^2$Instituto de Astronomía Te\'orica y Experimental (IATE), Laprida 854, C\'ordoba, Argentina\\
$^3$Department of Physics, Fudan University, 200438 Shanghai, China\\
$^4$Department of Physics and Astronomy, McMaster University, Hamilton, Ontario, Canada\\
$^5$Institute for Astronomy, University of Edinburgh, Royal Observatory, Edinburgh EH9 3HJ, United Kingdom\\
$^6$ Sorbonne Universit\'{e}, UPMC Univ Paris 06, and CNRS, UMR 7095, IAP, 98b bd Arago, F-75014, Paris, France\\
$^7$Departamento de F\'isica Te\'{o}rica, M\'{o}dulo 15, Facultad de Ciencias, Universidad Aut\'{o}noma de Madrid, 28049 Madrid, Spain\\
$^8$Centro de Investigaci\'{o}n Avanzada en F\'{\i}sica Fundamental (CIAFF), Universidad Aut\'{o}noma de Madrid, 28049 Madrid, Spain \\
$^{9}$International Centre for Radio Astronomy Research, The University of Western Australia, 35 Stirling Highway, Crawley, Western Australia 6009, Australia
}}

\date{Accepted 2020 April 16. Received 2020 April 14; in original form 2020 March 04}

\pubyear{2020}

\begin{document}
\label{firstpage}
\pagerange{\pageref{firstpage}--\pageref{lastpage}}
\maketitle

\begin{abstract}
Upcoming wide-field surveys are well-suited to studying the growth of galaxy clusters by tracing galaxy and gas accretion along cosmic filaments. We use hydrodynamic simulations of volumes surrounding 324 clusters from \threehundred project to develop a framework for identifying and characterising these filamentary structures, and associating galaxies with them.
We define 3-dimensional reference filament networks reaching $5R_{200}$ based on the underlying gas distribution and quantify their recovery using mock galaxy samples mimicking observations such as those of the WEAVE Wide-Field Cluster Survey. Since massive galaxies trace filaments, they are best recovered by mass-weighting galaxies or imposing a bright limit (e.g. $>L^*$) on their selection. 
We measure the transverse gas density profile of filaments, derive a characteristic filament radius of $\simeq0.7$--$1\hMpc$, and use this to assign galaxies to filaments. For different filament extraction methods we find that at $R>R_{200}$, $\sim15$--$20\%$ of galaxies with $M_*>3\times 10^9\Msun$ are in filaments, increasing to $\sim60\%$ for galaxies more massive than the Milky-Way. The fraction of galaxies in filaments is independent of cluster mass and dynamical state, and is a function of cluster-centric distance, increasing from $\sim13$\% at $5R_{200}$ to $\sim21$\% at $1.5R_{200}$.  
As a bridge to the design of observational studies, we measure the purity and completeness of different filament galaxy selection strategies. Encouragingly, the overall 3-dimensional filament networks and $\sim67$\% of the galaxies associated with them are recovered from 2-dimensional galaxy positions.
\newline 
\end{abstract}




\section{Introduction}
\label{sec:intro}

The matter distribution of the Universe follows a web-like structure, consisting of sheets, filaments, knots and voids, 
providing the environment in which galaxies form and evolve \citep{Bond1996}. According to current theories, this process is attributed to both external (e.g., interactions with the environment) and internal (e.g., galaxy stellar mass, feedback processes) physical mechanisms. However, galactic masses are highly dependent on their large-scale surrounding: intrinsic properties are intimately linked to their environment through their assembly process. 
Decoupling their complex interplay therefore requires the simultaneous exploration of the broadest possible range of masses as well as environments, defined both by their local density and the global Large Scale Structure (LSS) of the "cosmic web".

Filaments are ubiquitous in the Universe and account for 50-60\% of the matter in the Universe, but only $\sim6\%$ of the volume \citep{Cautun2014, Tempel2014}; (but see \citet{Cui2017, Martizzi2019, Cui2019a} for higher fractions). Cosmic filaments are elongated relatively high density structures of matter, tens of megaparsecs in length, that intersect at the location of galaxy clusters. They form through a gravitational collapse of matter along two principal axes: driven by gravity, baryonic gas traces the gradients of the dark matter distribution, shocks and winds up around multi-stream, vorticity-rich filaments \citep{ Codis2012, Laigle2014,Hahn2015,Kraljic2017}. This view of rich gas filaments feeding galaxy clusters based on simulations is firmly established and is now becoming available in gas observations \citep{Umehata2019}.

Not only do filaments play a key role in shaping galaxies,  
the cosmic web is also fundamentally connected to, and thus a probe of, cosmology.
According to current cosmological theories of structure formation, the early Universe was populated by small over-densities that grew through gravity. The web-like features of the large scale matter distribution were thus shaped by gravitational tidal forces.
Information about filaments is therefore embedded in the initial conditions of the Universe. 
In the highest density regions of the cosmic web, galaxy clusters formed hierarchically through the merging of smaller virialised halos. They continue to grow and assemble through a combination of smooth accretion and ingestion of smaller galaxy clusters and groups, which explains the complicated substructure that has been observed with increasing attention in the past decade \citep[e.g.,][]{Aguerri2010, Jaffe2016, Tempel2017}. 
The outskirts of galaxy clusters are therefore the points of contact that link the large scale cosmic web to the confined realms of cluster cores at their knots. They have emerged as one of the new frontiers and unique laboratories to study the mass assembly in the Universe as well as galaxy evolution in the context of global environment \citep{Walker2019}. However, much of the topology, geography and physics of cluster outskirts is fundamentally different from that of cluster cores -- and much less well understood.   
Identifying, mapping and characterising the low-contrast filamentary structures of the cosmic web provides invaluable information about galaxy formation, evolution and cosmology. In order to trace the impact of structure growth on the galaxy population, we must therefore consider galaxies in filaments out to and well beyond the cluster virial radius.

Observations of clusters show how fundamental the role of the environment is in shaping galaxies: morphology, colour, star formation rate (SFR), stellar age and AGN fraction correlate with both local galaxy density and location inside and outside clusters \citep{dressler80,blanton05, postman05, smith05,bamford09}.
During infall into clusters, the properties of the galaxies change. Quantification of these changes has mainly been focused on the end-point in the virialized regions of clusters. Nearby and intermediate-redshift cluster galaxy surveys (e.g., EDisCS, \citep{White2005}; WINGS, \citep{Fasano2006}; STAGES, \citep{Gray2009}; LoCuSS, \citep{smith10}) have studied the main properties (masses, morphologies, dynamics, star formation and AGN activity, scaling relations, etc) of the cluster population. As a result, much progress in our understanding of environmental mechanisms in the densest regions has been achieved.

Galaxies in over-dense environments are subject to astrophysical processes, including ram-pressure stripping of gas \citep[e.g.,][]{gunngott72, bahe17}, tidal effects  \citep[e.g.,][]{bekki98}, galaxy-galaxy interactions  \citep[e.g.,][]{naab07}, and mergers  \citep[e.g.,][]{hopkins08,Kaviraj2009}, that will disturb and remove their gas, ultimately resulting in the suppression of star formation \citep[e.g.,][]{delucia12,wetzel13}, and a change in morphologies and structures; see the reviews by \citet{Boselli2006} and \citet{Boselli2014}. As a consequence, we find many more red early-type (elliptical and S0) and fewer blue late-type (spiral and irregular) galaxies in clusters than in the field \citep{dressler97, Desai2007}. Accordingly, clusters have a lower fraction of star- forming galaxies \citep{Popesso2006} and cluster galaxies possess much less cold gas than field galaxies \citep{Cayatte1990}.
Hierarchical models of galaxy formation \citep[e.g.,][]{blumenthal84, Lucia2006} explain this observation with the argument that galaxies in the highest density peaks started forming stars and assembling mass earlier. In essence, they have a head start \citep{Bond1991}, so one would expect that galaxies in high-density environments preferentially host older stellar populations. Simultaneously, galaxies forming in high-density environments will have had more time to experience the external influence of their local environment. 

%
%
It is important to consider that infalling galaxies account for approximately half of a cluster population, and so contribute to a growth in cluster mass of 100\% by today \citep{dressler13, McGee2009}. 
Therefore, a significant fraction of cluster galaxies has been environmentally affected long before they reach the cluster centre, a concept termed "pre-processing". 
In fact, the transition from "field-like" to "cluster-like" populations starts to occur beyond 1--2 virial radii from the cluster centre, experiencing pre-processing in outskirt evironments  \citep[e.g.,][]{Haines2015, Haines2018, Kuchner2017, Bianconi2017}. It is clear that we need to extend our environment considerations to the idea that a galaxy has experienced a variety of environments over its lifetime as part of the cosmic web and infall region of clusters.

Within this context, several recent photometric and spectroscopic surveys have focused on the contribution of the global structure features of the cosmic web (knots, sheets, filaments and voids) to galaxy evolution.
They report that galaxy colour, mass, morphology, fraction of passive and star forming galaxies and sSFR vary with distance to filaments in the cosmic web in the sense that galaxies nearer filaments are redder, more massive, have reduced star formation rates and tend to be elliptical \citep{Alpaslan2016, Laigle2017,Kraljic2017, Kuutma2017, Sarron2019, Liu2019}. Contrariwise, other observations find some evidence for intriguing HI enhancements near filaments of the cosmic web \citep{Kleiner2016, Vulcani2019}, suggesting a "cosmic web enhancement". 
Though the controversy is not solved, this suggests that the multi-stream region of the large scale structure does have a secondary effect (besides the local environment) and that galaxies accreted by clusters indeed become affected well before they reach the cluster centre.

In response to these challenges, future surveys 
will explore the filamentary structures far beyond the virial radius of clusters as important sites of galaxy evolution. Surveys like the WEAVE wide-field cluster survey or the 4MOST cluster survey \citep{Finoguenov2019} are designed to chart and characterise cluster environments from the densest cluster cores to the lower-density filamentary infall regions that surround them and will therefore be able to shed light on pre-processing mechanisms in outskirts. 
Given the complexity of identifying consistently and robustly galaxies belonging to these structures \citep[e.g.,][]{Martinez2015, Laigle2017,Malavasi2017, Kraljic2017, Kraljic2018, Sarron2019}, observations alone are not enough.
It is imperative that realistic simulations are used to develop and test reliable structure-finding methods and to characterise their robustness and uncertainties. Simulations are thus essential for the planning and design of the targeting strategy of future surveys and will play a crucial tool in interpreting their results. 

In this paper, we summarize our intention to prepare for the upcoming WEAVE Wide-Field Cluster Survey (WWFCS, Kuchner et al in prep.). WEAVE (WHT Enhanced Area Velocity Explorer, \citet{Dalton2012, Balcells2010}) is a new multi-object survey spectrograph for the 4.2-m William Herschel Telescope (WHT). WWFCS will make use of the instrument's two-degree diameter field of view multi-object spectrograh (MOS) with up to 1000 targets in a single exposure \citep{Sayede2014}. The survey is designed to map, characterize and study infall regions of 16--20 galaxy clusters out to $5 \times R_{200}$ with an unprecedented number of structure members down to a mass limit of $M_*=10^9 \Msun$. 
Here, we focus on the preparation steps using simulations of clusters to develop techniques to 1) optimally find filaments and 2) associate galaxies to them. We use simulations from \threehundred project, which has completed resimulations of the 324 most massive galaxy clusters and their surrounding environment from the MultiDark $1~h^{-1}\rm{Gpc}$ simulation (MDPL2), to test the robustness and reliability of detecting filaments in an observational framework.
In Section \ref{sec:simulations}, we introduce the simulations and summarise the filament extraction using smoothed gas particles.
We also discuss preferred alignments of gas filament and their thickness.
We then move towards observations (Sec. \ref{sec:observations}) and identify filaments using mock galaxies based on well-founded detection limits. To assess their reliability, we investigate the effects of going from idealised gas to mock galaxies and from 3D to projected 2D mock galaxy distributions. We then discuss how galaxies associate to filaments, and report an accumulation of galaxies in filaments closer to the cluster. Finally, an evaluation of the performance in several realistic cases aims to provide practical decision-making support for observations. We summarise our findings in Section \ref{sec:conclusion}.


\section{Simulations}
\label{sec:simulations}

\subsection{\threehundred cluster project}
\label{subsec:the300}
\begin{figure}
   \centering
   \includegraphics[width=9cm]{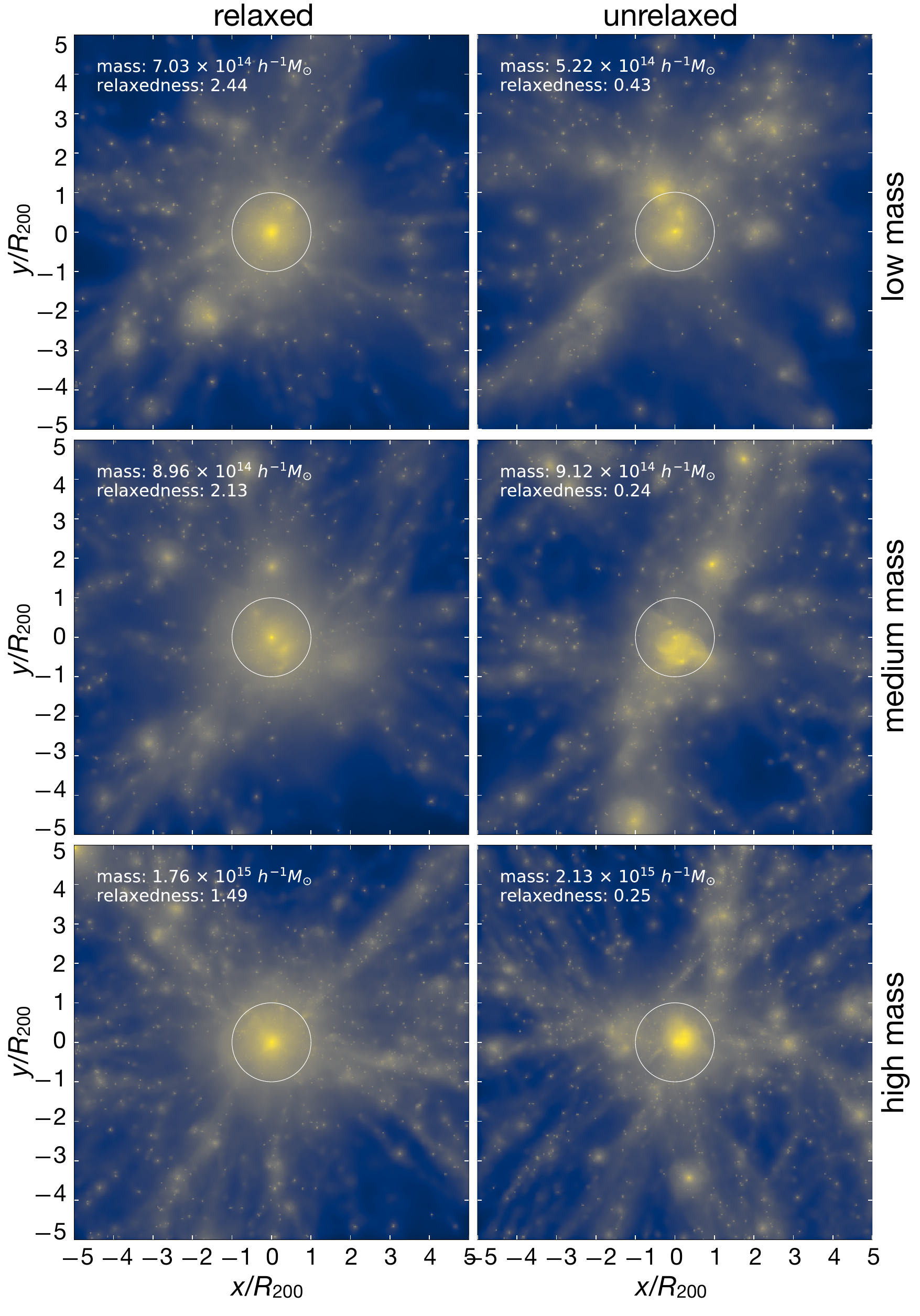}
    \caption{Gallery of six galaxy cluster volumes from \threehundred project at $z=0$. Shown are projected gas distributions within 5$R_{200}$ for a range of mass and "relaxedness", a measure for the dynamical state of the central region of a cluster. It is quantified by combining the fraction of mass in sub-halos, the centre-of-mass offset and the virial ratio (see section \ref{subsec:relaxedness} and figure \ref{fig:relaxedness}). The circles indicates $R_{200}$ of each cluster; mass and relaxedness values of the examples are printed in each left upper corner. Clusters with high relaxedness values are more relaxed.}
   \label{fig:gallery}
\end{figure}

In this paper we use 324 simulations of massive clusters and their surrounding environment from \threehundred project\footnote{\url{https://the300-project.org}} \citep[][Ansafari et al. in prep]{Cui2018, Wang2018a, Mostoghiu2018, Arthur2019}.
The simulations are re-simulated zoom regions of the dark-matter-only MDPL2, Multi-Dark 1Gpc/h simulation \citep{Klypin2016}. MDPL2 uses {\em Planck} cosmology ($\Omega_{M}$ = 0.307, $\Omega_{B}$ = 0.048, $\Omega_{\Lambda}$ = 0.693, $h$ = 0.678, $\sigma_{8}$ = 0.823, $n_{s}$ = 0.96) and $3840^{3}$ dark matter particles per co-moving $1~h^{-1}\rm{Gpc}$ box.
\threehundred project then selected the 324 most massive galaxy clusters at $z = 0$, followed them back to initial conditions and re-simulated them with higher resolution in regions of radius $15\hMpc$. 
These simulations use the \gadgetx full-physics galaxy formation code incorporating star formation and feedback from both SNe and AGN. They were modelled using a Smoothed Particle Hydrodynamics (SPH) algorithm with a subgrid physics scheme \citep{Beck2015} to follow the gas component's evolution with a combined mass resolution of $m_{\rm{DM}} + m_{\rm{gas}} = 1.5 \times 10^9~h^{-1} \Msun$ \citep[see][for details]{Cui2018}. 
For our purpose, we make use of the full physics information in dark matter, gas and halo distributions in the simulated boxes.

Figure \ref{fig:gallery} shows six examples of clusters from \threehundred project at $z=0$. Each resimulation contains a central cluster within a sphere extending to $15\hMpc$, covering the entire cluster infall region and associated filaments. On average, the high resolution regions reach well beyond 5$R_{200}$ of the cluster's dark matter halo, where $R_{200}$ represents the radius within which the mean density is 200 times the critical density of the Universe. 
129 snapshots from $z = 16.98$ to $z = 0$ are available with a mass resolution of $10^8 \Msun$ each. Their masses range from \mbox{$M_{200} = 6.08 \times 10^{14}~h^{-1}\Msun$} to $M_{200} = 2.62 \times 10^{15}~h^{-1}\Msun$.
The mass-complete sample covers the full range of cluster dynamical states, including both relaxed and currently merging objects (see Sec. \ref{subsec:relaxedness}). Figure \ref{fig:gallery} highlights the diversity of clusters, showing examples of relaxed and unrelaxed clusters from low, medium to high masses. 

In an extensive comparison project, the {\em nIFTy cluster comparison project} \citep{Sembolini2016,Sembolini2016a, Elahi2016, Cui2016a, Arthur2016a,Power2019}, a progenitor project of The ThreeHundred project, authors compared ten different simulation codes. These were run on one example galaxy cluster that was simulated both using dark matter only and including baryonic physics. In the case of the dark-matter-only cluster, the different simulation codes perform in agreement with each other. When baryon models were taken into account, only the overall cluster properties (e.g., $M_{200}$) were recovered in the different simulation codes. On small scales, however, the test revealed significant discrepancies. 
Since one cluster does not provide enough statistics to compare with observations or to distinguish the various models, \threehundred project was established: it encompasses 324 clusters, each with baryon models \textsc{Gadget-MUSIC}, \gadgetx (used in this work), and \textsc{GIZMO} (Cui et al. in prep.), as well as three semi-analytical models which are based on MultiDark-Galaxies \citep{Knebe2017}. These have been used to populate the entire MDPL2 simulation volume with galaxies, thus generating realistic background and foreground samples, as well as full light-cones.
In addition, the hydrodynamic simulations provide true six-dimensional phase space views of structures and substructures \citep{Arthur2019}, where observations are limited to line-of-sight views. 
More information, as well as visualisations and movies of the 324 resimulated clusters at different epochs can be found at:
http://www.mockingastrophysics.org and http://music.ft.uam.es/videos/music-planck.

\subsubsection{Data products: halo catalogues}
\label{subsec:data_products}

The analysis is based on halo catalogues for all 324 clusters extracted with the \Amiga\ Halo Finder \citep[ \textsc{AHF};][]{Gill2004, Knollmann2009, Knebe2011} that includes gas and stars in the halo finding process self-consistently. In a nutshell, \textsc{AHF} finds the prospective halo centres trough following density contour levels from high to background densities in trees of nested grids, then collects particles that are possibly bound to the centre, removes the unbound particles, and calculates the halo properties (i.e., halo and stellar masses, virial radii, geometry, density profile, velocity dispersion, peculiar velocities, rotation curve). All halo properties are based on all particles inside the halo,
i.e., dark matter, gas, and (if available) star particles, inside a sphere of radius $R_{200}$ that defines the halo edge. This is at a distance of the farthest gravitationally bound particle inside a "truncation radius", and the point where the density profile of bound particles drops below the virial overdensity threshold as given by cosmology and redshift.
\textsc{AHF} organises the output in a tree structure with information about hosts, subhalos, sub-subhalos.  Far-UV to sub-mm luminosities are calculated from the stellar population synthesis code STARDUST \citep{Devriendt1999} providing a reference for standard photometric bands such as SDSS's for optical scaling relations \citep[see][for details]{Cui2018}. 

In this analysis, we are using the following properties:
\begin{itemize}
\item Halo: \textsc{AHF} classifies halos as objects made of dark matter and baryonic particles. We do not make any specific distinction between halo and subhalo in this paper.

\item Cluster halo: The most massive halo in each re-simulated volume at $z = 0$, which is also the centre of the simulation box. We also identify the second most-massive halo, SMH.

\item $R_{200}$: the radius of a sphere where the mean density is 200 times the critical density of the Universe.

\item $M_{200}$: The mass enclosed within a sphere of radius $R_{200}$, given in $\Msun~h^{-1}$.
\end{itemize}

\subsubsection{Dynamical relaxation of clusters}
\label{subsec:relaxedness}
\begin{figure}
   \centering
   \includegraphics[width=\columnwidth]{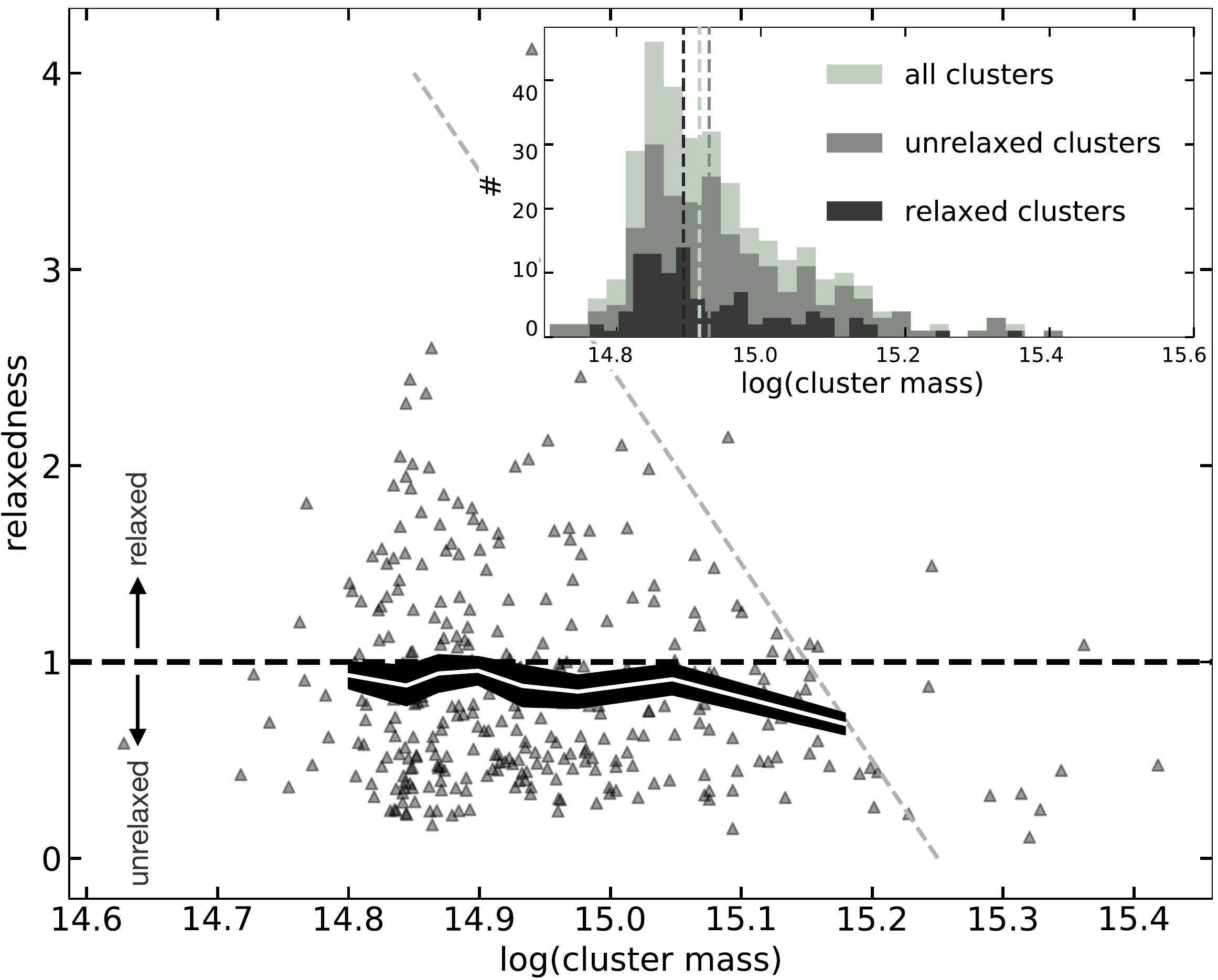}
    \caption{Galaxy cluster mass as a function of relaxedness for 324 clusters in \threehundred simulations. Clusters are divided into unrelaxed ($R$<1) and relaxed ($R$>1) populations (see text for description). High mass clusters are usually unrelaxed; they are dynamically active e.g., through accreting matter from their surroundings. Low mass clusters show a wide range of $R$. The black error bands shows the average of all simulated clusters in our sample and is neither dynamically active nor relaxed.  The diagonal dash-line marks the approximate location of the envelope of the point distribution discussed in the text. The insert shows the histograms of all clusters (light shade), unrelaxed clusters (medium shade) and relaxed clusters (dark shade). Dashed lines in the insert indicate the median values and show the preference for low (high) mass clusters to be relaxed (unrelaxed).}
   \label{fig:relaxedness}
\end{figure}

Accretion physics leaves characteristic tracers in the infall regions of galaxy clusters. Ongoing accretion is accompanied by signatures of dynamical activity typical for unrelaxed clusters. In order to identify whether the accretion of matter via filaments is correlated to the dynamical state of a cluster, we categorise the clusters by their "relaxedness". To determine the dynamical state of a cluster, \citet{Cui2018} introduces three parameters: 
\begin{itemize}
\item the virial ratio, a measure of how virialized the clusters is, defined as $\eta = (2T - E_{\rm{s}})/|W|$, where T is the total kinetic energy, $E_{\rm{s}}$ is the energy from surface pressure, and $W$ is the total potential energy,
\item the centre-of-mass offset from its point of highest density (which typically coincides with the brightest cluster galaxy), defined as $\Delta_r = |R_{\rm cm} -R_{\rm c}|/R_{\rm 200}$, where $R_{\rm cm}$ is the centre-of-mass within a cluster radius of $R_{\rm 200}$ and $R_{\rm c}$ is the centre of the cluster defined as the maximum density peak of the halo,
\item the fraction of cluster mass in subhalos $f_{\rm s} = \Sigma M_{\rm {sub}}/M_{\rm {200}}$, where $M_{\rm {sub}}$ is the mass of each subhalo.
\end{itemize}
A combination of these three parameters defines the dynamical state of a cluster as either "relaxed" or "unrelaxed". In this framework, a given cluster is "relaxed" if it satisfies, $0.85 < \eta < 1.15$, $\Delta_{\rm {r}} < 0.04$ and $f_{\rm s} < 0.1$. This means that we expect a relaxed cluster to have a low fraction of mass in sub-halos, low centre-of-mass offset and a virial ratio equal to 1. A "maximally relaxed" cluster thus has values of $\eta = 1,\, \Delta_{\rm {r}} = 0$ and $f_{\rm {s}} = 0$, and unrelaxed clusters begin at $|\eta -1| = 0.15,\, \Delta_{\rm {r}} = 0.04$ and $f_{\rm {s}} = 0.1$.
We combined these into one general parameter $R$ in the following way: 
\begin{equation}
R = 1 / [ (1/3) * ( ((\eta-1) / 0.15)^2 + (\Delta_{\rm r} / 0.04)^2 + (f_{\rm s} / 0.1)^2 ) ]^{0.5}
\end{equation}

We use this parameter $R$ to describe how relaxed a cluster is \citep[see also][]{Haggar2020}. Clusters with a greater value for $R$ are more relaxed, and $R = 1$ is roughly equivalent to the division between relaxed and unrelaxed clusters used in \citet{Cui2018}.

In Figure \ref{fig:relaxedness}, we plot the dynamical state of the clusters at redshift 0, i.e., its "relaxedness", versus the cluster mass. The solid black fit line shows a relatively flat rolling average value over all masses. The average relaxedness of all clusters in our sample is roughly 1, and thus neither especially relaxed nor unrelaxed.
The dashed line in the main panel indicates an "envelope" suggesting that the most massive clusters tend to be currently in un-relaxed states. Its purpose is purely to guide the eye and is drawn by hand. This is an indication that the most massive clusters are still growing today, in complex ways that result in complicated substructure and centre-of-mass offsets. In addition, the most massive clusters are located in high density regions of space, and thus have a higher likelihood of accreting matter from their surrounding dynamic cluster environment -- resulting in unrelaxed dynamical states.
Low mass clusters spread over a range of dynamical states, from completely unrelaxed to relaxed. Isolated low mass clusters may have grown a long time ago and have had time to relax since then. Alternatively, they could have started accreting mass from the cosmic web only recently. 
Thus, clusters of all masses can be unrelaxed. Our checks for resolution effects rule out the possibility that the resolution of the simulations are the cause of the described scenario.
The insert shows histograms of all (light shade), relaxed (dark shade) and unrelaxed (intermediate shade) clusters separately. Throughout most of the mass range of \threehundred simulations, unrelaxed clusters consistently make up about two thirds of the total cluster distribution. The dashed lines indicate the median values and clearly show that unrelaxed clusters preferentially have lower masses than relaxed clusters.  

\begin{figure}
   \centering
   \includegraphics[width=\columnwidth]{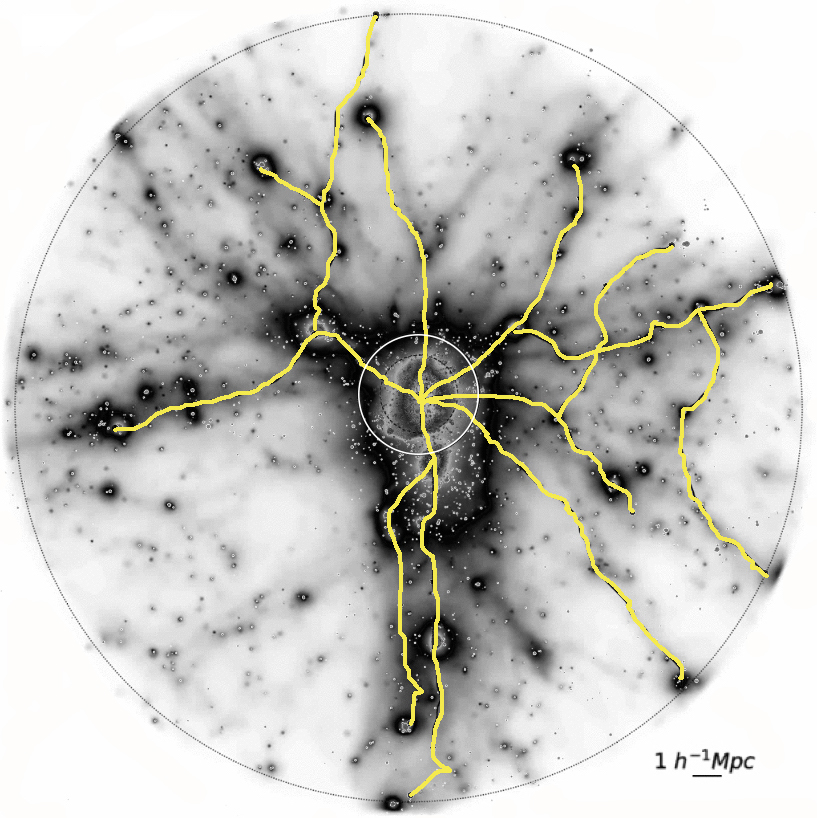}
    \caption{Example filament network (yellow lines) of the central node of cluster 0001 from \threehundred project, based on the geometric three-dimensional ridge extractor \disperse\ (see text for details). The figure demonstrates the filament extraction of our reference filament network using smoothed gas particles. Seen is the projected gas distribution at $z = 0$ within a $15\hMpc$ sphere of the central cluster; $R_{200}$ is shown as a white circle.}
   \label{fig:disperse_DM}
\end{figure}
\subsection{Filament finding}
\label{sec:finding_filaments}
This work focuses on quantifying the bias of using galaxies as tracers of cosmic filaments in cluster outskirts. Filaments are identified from a density field. In our case, this refers to either the number density of simulated gas particles, which we use to extract a three-dimensional reference skeleton, or to the number density of mock galaxies, i.e., halos matched to observable galaxies. 
The accuracy of the reconstruction of the filament network depends on the sampling of the data set. 
We therefore use the capacity of \threehundred simulations to compare filament reconstructions from the underlying idealised case with a realistic setup of future cluster outskirt observations (Sec. \ref{sec:observations}). 

\subsubsection{Cosmic filament reconstruction with \disperse}
\label{subsec:disperse}

Our reconstruction of filamentary networks around clusters is based on the DIScrete PERsistent Structure Extractor (\disperse\, \citep{Sousbie2011}).
The algorithm is based on the discrete Morse theory and theory of persistence, and is explained in \citet{Sousbie2011}. In short, the software utilises a discrete distribution of points -- in our case coordinates of halos or gas particles -- to reconstruct the volume as cells, faces, edges and vertices. The density of this distribution is estimated form the Delaunay tessellation of the points.
In practice this means that the Delaunay Tessellation Field Esimator \citep[DTFE;][]{Schaap2000, Cautun2011} calculates the density around each vertex of the Delaunay complex. The algorithm does this by first computing a triangulation on the field, then the density in each cell is computed as the inverse area of the cell.
To calculate filaments and nodes (i.e., peaks) from this density field, \disperse\ extracts the critical points, i.e., points where the gradient is null of the density field like maxima, minima and saddle points, and links them along ridges. 

The connections between the critical points are field lines tangent to the gradient field in every point. 
\disperse\ computes a series of individual small \textit{segments} that define \textit{ridges} which link topological saddle points to nodes and together they form a skeleton that identifies the filamentary network \citep{Pogosyan2009} in our simulation. These are arcs, linking critical points; in 3D, maxima are critical points of order 3 (2 in 2D) and saddle points are critical points of order 2 or 1 (1 in 2D).
Thus, each filament is constructed as a set of segments that join nodes to saddle points or bifurcations. 
Persistence quantifies the ratio of the density value, i.e., the density contrast, of a pair of specific critical points like node to saddle points. The persistence level is therefore a measure of the significance of topological connections between critical points (comparable to a minimal signal-to-noise ratio) and is usually expressed as a number of standard deviation $\sigma$.
Because the cosmic web and thus the filament network is multiscale, the persistence threshold is crucial for the definition and robustness of filaments: choosing the persistence allows to filter noisy structures. A larger persistence threshold tends to isolate the topologically most robust filaments. 

Filament extraction can be done in 3D and in 2D, directly using discrete data sets of coordinates, regardless of scale or persistence levels. This means that \disperse\ is equally applicable for the feature extraction based on a density field of gas particles of \threehundred simulations as it is based on observations of galaxies. 
Several authors have recently shown how \disperse\ can be used to trace the cosmic web on large scales using simulations \citep[e.g.,][]{Dubois2014} and observations \citep{Malavasi2017, Kraljic2017, Laigle2017}, both in (projected) 2D and 3D.
In these examples, the maxima (e.g., galaxy clusters and groups) are linked by filaments of several Mpc to several tens of Mpc in length, depending on the sampling. 
On smaller scales like in the case of a \threehundred simulation box, with only one cluster and its surrounding infall region, the saddle-point along the filament linking this cluster (node) with the next might be outside the simulation box (field-of-view). In the presented case of a simulation box with $15\hMpc$ co-moving length, filaments may only be 2--3 times longer than they are thick. However, because \disperse\ is scale-free, it can extract features independent of their scale, largely depending on the persistence threshold that the user chooses. For a comprehensive comparison between a number of available filament finders, including \disperse, and the different methods they employ, we advice the reader to refer to \citep{Libeskind2017}.


\subsubsection{Filament extraction using smoothed gas particles}
\label{subsec:fil_gas}

We define simulated gas filaments as the reference frame for our assessment. We therefore first identify the 3-dimensional filamentary network of the underlying gas distribution in each of the re-simulated volumes using \disperse's topological method. We choose to use the distribution of gas particles rather than dark matter particles because, as an observable property, gas may be accessible for future surveys. 
Note, however, that while the distribution of gas follows dark matter -- and thus alludes to the underlying distribution of dark matter -- some variation between dark matter and gas skeletons are expected.  
Because our aim is to use gas filaments as the benchmark for galaxy filaments, we chose persistence levels that lead to filaments with high contrasts. Note that the simulations would give access to many more lower density gas filaments (tendrils) that are inaccessible to the observational constraints we use in this paper and thus irrelevant for the present case \citep[see][for a detailed discussion]{Welker2019}.

To find gas filaments, we first bin the gas particles in a 30 Mpc-wide 3-dimensional grid with a resolution of size $150~h^{-1}$ co-moving kpc using a cloud-in-cell algorithm. The grid is gaussian-smoothed over eight times the pixel length. This method allows to focus on cosmic filaments that connect groups and clusters rather than thin filaments e.g., between large satellites. We then extract the filament network using an absolute persistence cut of 0.2. Expressed in standard deviations of a minimal signal-to-noise ratio, this translates to a $5\sigma$ persistence threshold. 
This value was chosen to ensure that cluster centres and massive groups are detected as nodes, and filaments connected to the main halo terminate in saddle points\footnote{Note that the choice of persistence parameter depends on the science question. Lower thresholds would reveal a wealth of thinner filaments (tendrils) with more nodes and saddle points.}
Subsequently, we cleaned and simplified the \disperse\ outputs for our purposes by matching the ends of segments and tracing the matches from each saddle point. We treat each node as owning its own network, connected by saddle points at the lowest density. 
Figure \ref{fig:disperse_DM} shows the filamentary network associated with the central object of one of the clusters  in \threehundred database. The background shading shows the projected gas density; the white inner circle marks $R_{200}$ of the cluster. Most branches terminate within the sphere of $15\hMpc$ radius encompassing the cluster, shown as the outer grey circle. Within this region, the full treatment of the physics ensures a realistic and suitable representation of the filamentary structure around massive clusters.

\subsubsection{Stability of filament networks over time}
\label{subsec:stability}

\begin{figure}
   \centering
   \includegraphics[width=\columnwidth]{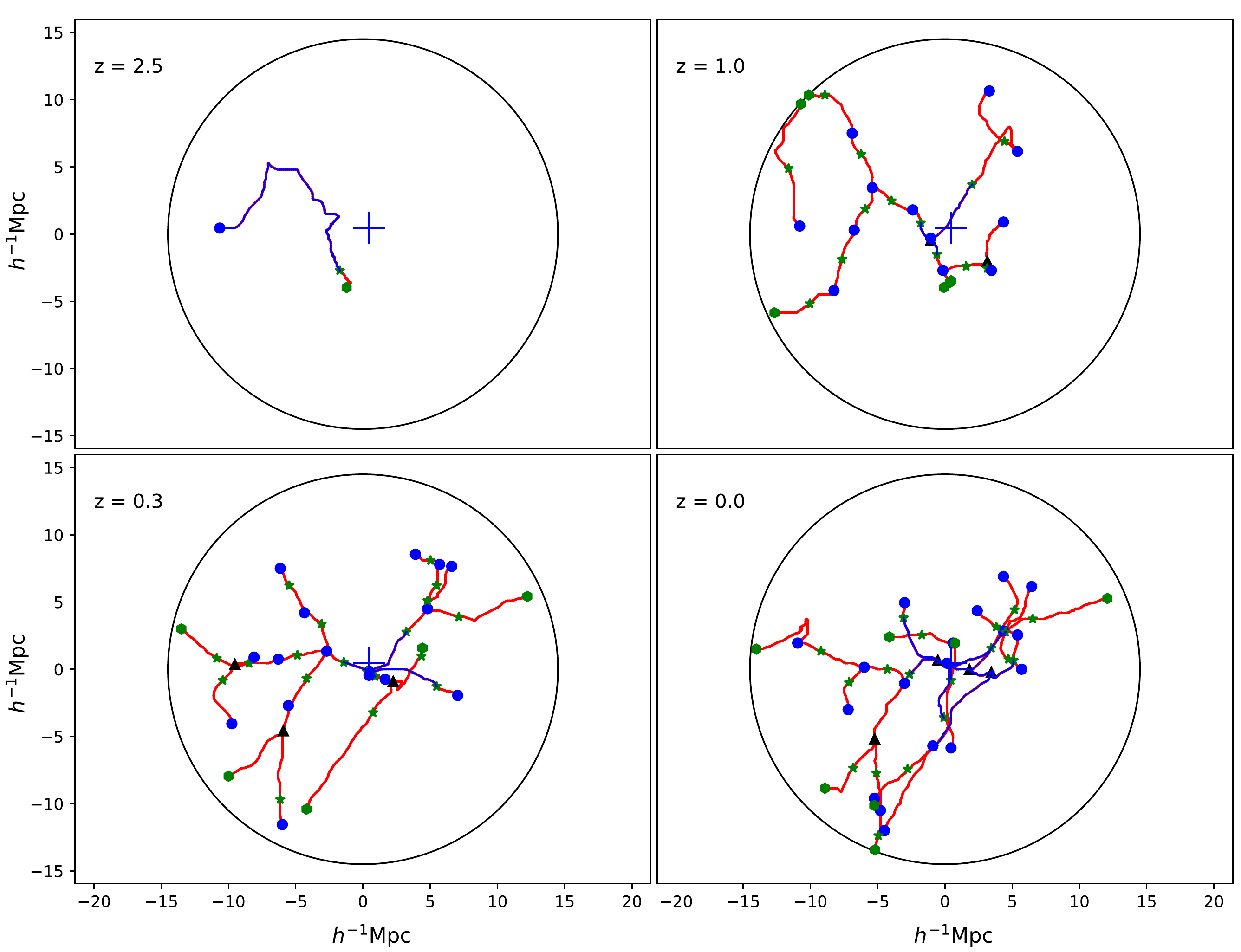}
    \caption{The figure shows the filament network of one example cluster using gas particles at four time steps from $z=2.5$ to $z=0.0$. Filaments were extracted from each frame independently and are shown here to demonstrate the stability of the filament finding with \disperse. Red lines indicate skeletons based on smoothed gas particles, blue dots mark nodes, green stars saddle points and black triangles bifurcations. Filaments coloured in blue further indicate the network connected to the central node of the cluster. Filaments leave the volume at a radius of $15\hMpc$.}
   \label{fig:stability}
\end{figure}

One way to further verify the reliability of the filament networks is to examine their stability over time. 
\threehundred project provides 129 snapshots between $z=17$ and $z=0$ for each cluster. We processed all time steps up to $z=2.5$ in the manner described above, thus retrieving the filamentary history of each cluster as an evolutionary stack. 
Fig. \ref{fig:stability} shows four example snapshots at different time steps, at $z=2.5,z=1, z=0.3$ and $z=0$, which is a fair representation of the entire evolution sequence we qualitatively investigated. 
Even though there is no connection in the algorithm between one output and the next, the nodes and filamentary network controlled by the central object remain smooth and stable  when we join the sequence of outputs. 
The sequence suggests that the cluster and its filamentary structure around it evolves over time.
While there is significant expansion of the volume between redshifts one and zero, we see the networks become more complex with time: The networks condense as the volume collapses and more particles fall onto the middle, while they continue to expand further out. This explains why more filaments appear at later epochs. 

For our purposes in this paper, we use this qualitative assessment solely as a further indication for the reliability of the performance of the filament finding with \disperse. All results that follow in this paper are based on simulations at $z=0$.

\subsection{Filament characteristics}
\label{subsec:characteristics}

\renewcommand{\thesubfigure}{\roman{subfigure}}
\begin{figure*}
  \centering
  \subfloat[Stacked filament networks]{\includegraphics[width=0.33\textwidth]{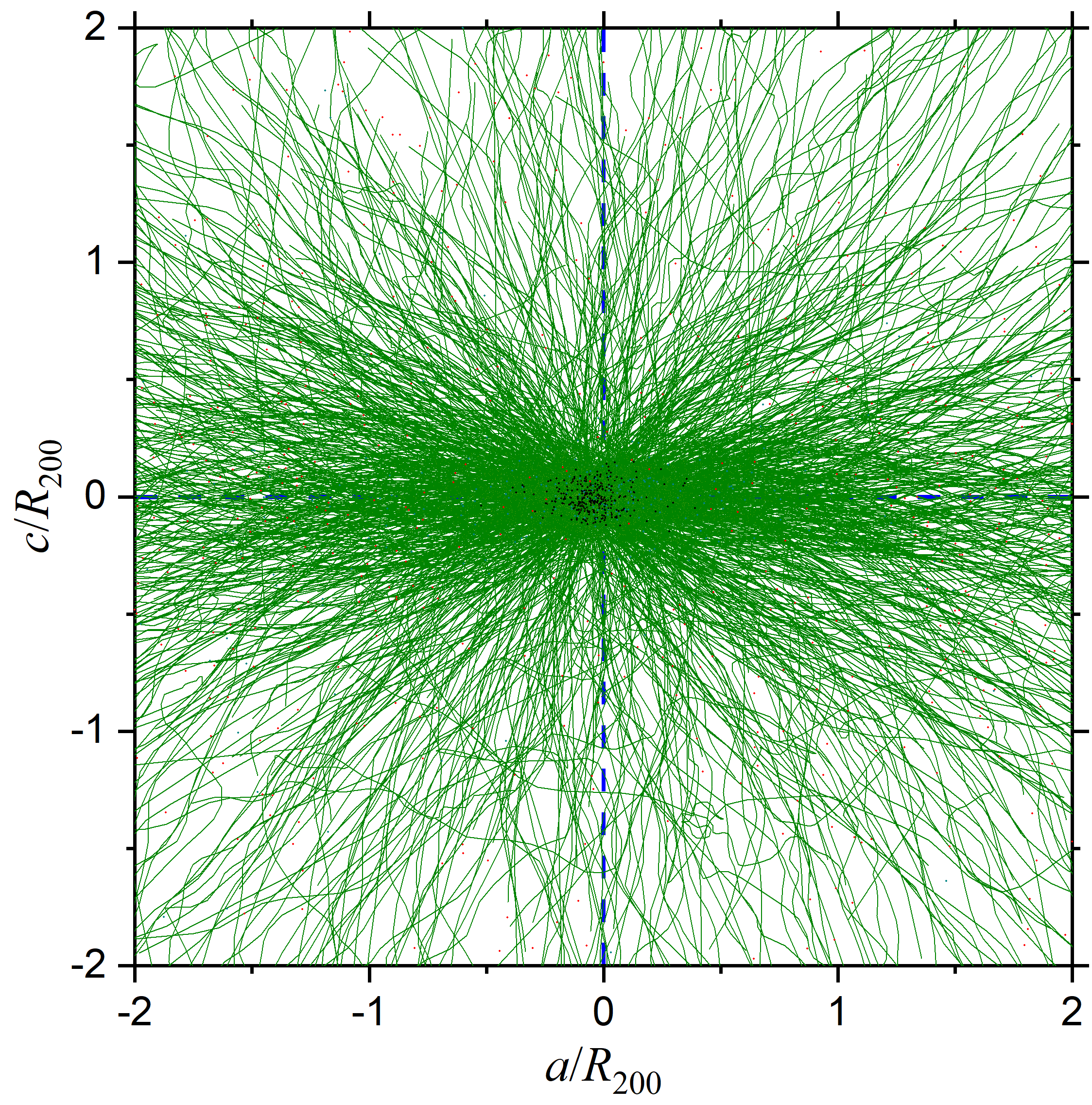}\label{fig:principle_filaments}}
  \hfill
  \subfloat[Angular distribution]{\includegraphics[width=0.33\textwidth]{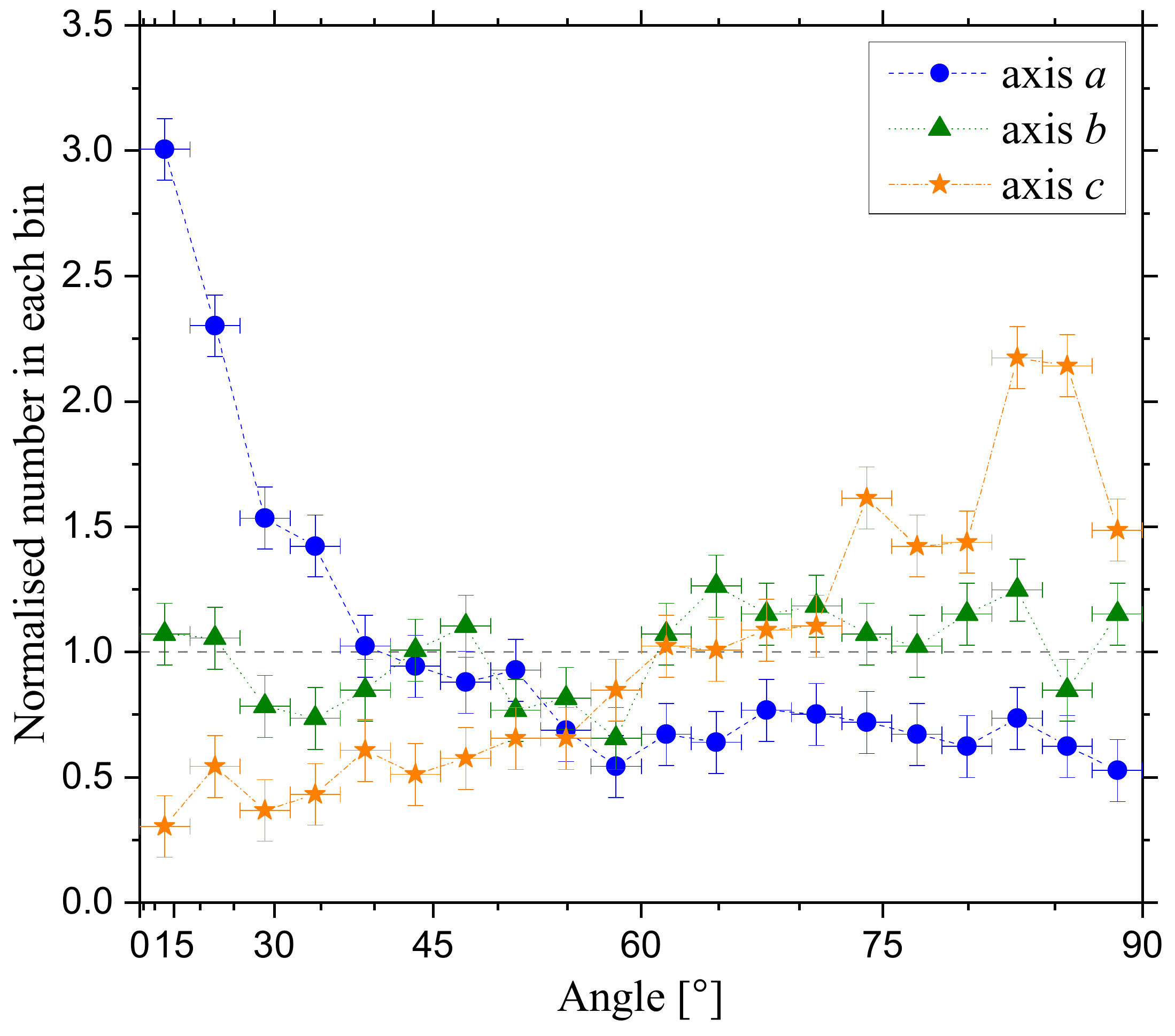}\label{fig:histo_stack}}
\hfill
  \subfloat[Second most massive halo]{\includegraphics[width=0.33\textwidth]{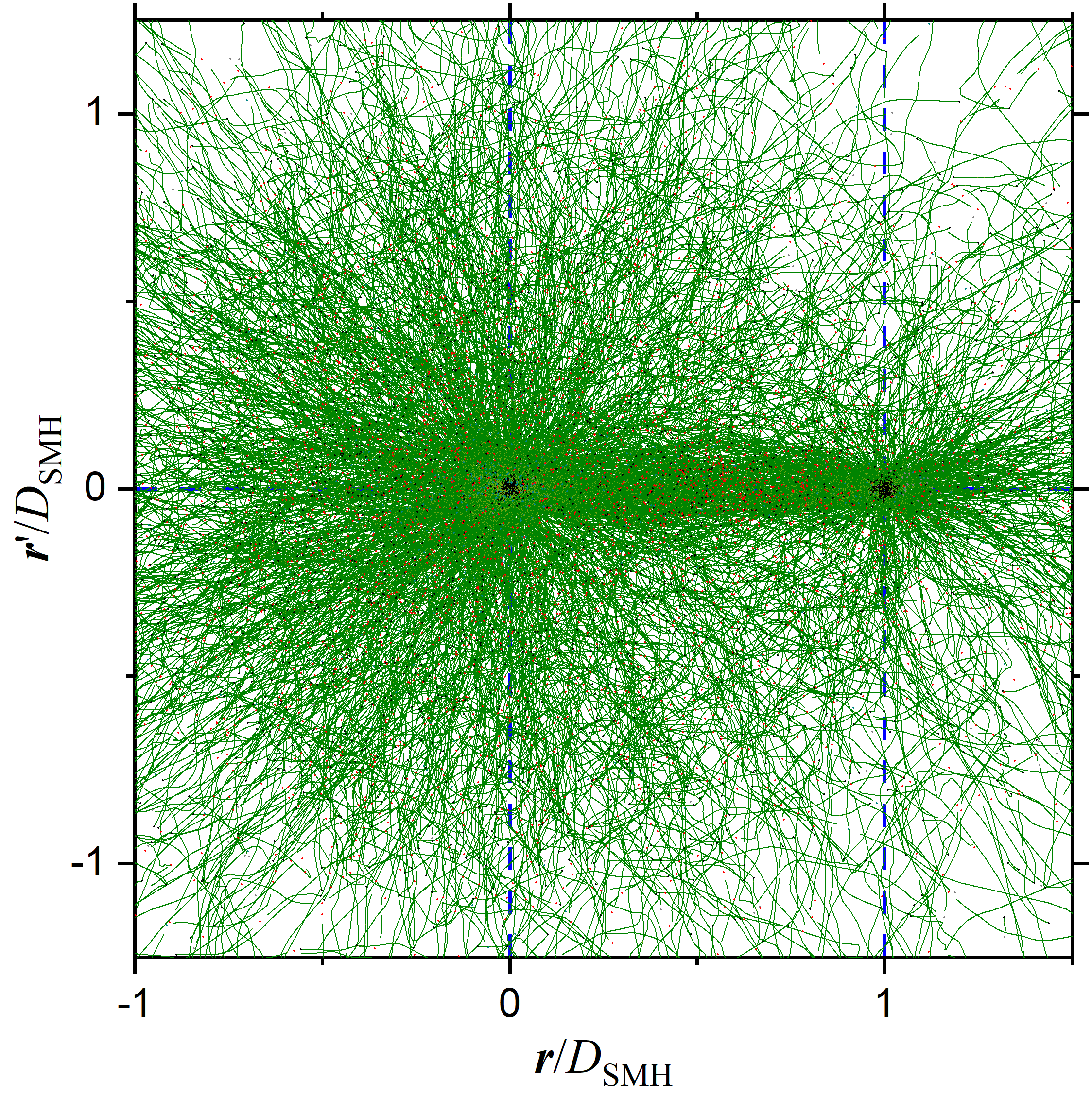}\label{fig:secondhalo}}
  \caption{Filament networks connected to the most massive (central) halo align with the principle axis of the main cluster halo. (i): Stacked filament networks extracted from \disperse\ on gas particles, connected to the main node, stacked according to the major axis $a$ of the clusters and normalised by $R_{200}$. In this figure, we chose to show the stacked network along the ac-plane. Only main filaments that ended at saddle points are included in this plot. (ii): Distribution of the angles of filaments when leaving a sphere of radius $R_{200}$ and the principal axes $a$ (blue points), $b$ (green triangles), $c$ (orange stars) that describe the shape of the main halo. (iii): Alignment of filaments with the second massive halo (SMH) in the simulation box. The coordinate system is rotated and re-scaled to $D_{\rm{SMH}}$, the distance between the main halo and the second massive halo. \textbf{r} is the position vector from the main halo to the second most massive halo, 
  $\textbf{r}'$ is a vector perpendicular to \textbf{r}. 
  }
\end{figure*}
\subsubsection{Filaments align with the shape of the central halo}
\label{subsec:alignment}

The filamentary envelope of clusters mark non-spherical accretion of material. Ultimately, this fuels the hierarchical assembly of massive structures. The preferred directions of accretion influence the shape and angular momentum of halos, also responsible for large scale alignments \citep{Aragon-Calvo2007, Hahn2007}. 
In order to investigate how matter in the Universe is accreted onto clusters, we test the alignment of filaments extracted from gas particles with the overall shape of the main dark matter halo -- a proxy for the shape of the galaxy cluster as a whole. This could reveal preferred inflow directions that are responsible for building the cluster. 
We investigate correlations between the alignment of filaments to the shape (geometrical axes and elongation) of the central halo and the influence of the second most massive halo in the simulation box.

We find that filaments connected to the main halo preferentially align with the major axis of this halo. We characterise the shape of each simulated halo by three axes ($a, b, c$ from major to minor in our illustrations), that describe their triaxial nature. We extract these measurements from the \Amiga\ Halo Finder \textsc{AHF} results of the dark matter particles (see Sec. \ref{subsec:data_products}). Each cluster simulation box is dominated by a central halo that typically accounts for $\sim$90\% of the overall cluster mass. We therefore consider this halo a valid approximation for the entire cluster and measure alignments of filaments with respect to the axes of this main halo. 
For our analysis, we rotated each cluster to align on a common axis and stacked all networks of each principal node, normalized by $R_{200}$. Figure \ref{fig:principle_filaments} visualizes this stacking procedure projected onto a 2D plane and demonstrates the preferred alignment of filament with the principal axis, indicated by $a$.

To quantify this result, we follow the procedure reported in \citet{Veena2018}. For each filament of the main node, we measure the angle at which a filament exits a sphere of $R_{200}$ radius. By comparing this angle with angles measured from a random distribution of filaments (dashed horizontal line in Fig. \ref{fig:histo_stack}) allows us to quantify the significance of the alignment.  
%
This is shown in Figure \ref{fig:histo_stack}
: the blue histogram has a sharp peak around 0\textdegree, while the histogram showing alignments with the minor axis ($c$, in orange) consequently counteracts this at 90\textdegree. This is of course explained by the fact that a,b, and c are not independent, rather, they are orthogonal. Any vector that is parallel to one of them is inevitably perpendicular to the others. The finding supports the view that filaments are aligned with the shape of galaxy clusters in the inner region, in line with previous studies \citep{Hahn2007, Zhang2009, Libeskind2012, Veena2018}. 

Filaments further align more prominently in elongated clusters. For this investigation, we define a halo elongation coefficient $\delta_{\rm{el}}$ as the standard deviation: 

\begin{equation}
    \delta_{\rm{el}} \equiv \sqrt{\frac{(|a|-\bar{x})^2 + (|b|-\bar{x})^2 + (|c|-\bar{x})^2}{3}}
\end{equation}

We divide the sample of 324 clusters into 3 groups of equal size according to their central most massive halo's elongation $\delta_{el}$ and find that filaments align more strongly with the major axis in elongated clusters. In strongly elongated clusters ($\delta_{\rm{el}}>0.145$), 38.5\% of all filaments leave $R_{200}$ within an angle smaller than 30\textdegree\, to the major axis. In clusters with medium elongation ($0.13<\delta_{\rm{el}}<0.145$), the percentages decreases to 32.3\% and for the least elongated bin ($\delta_{\rm{el}}<0.13$), only 26.3\% of filaments leave within 30\textdegree\, of the major axis. 
The alignment effect is especially striking close to the central halo and weakens as we move further away from $R_{200}$, which we tested by measuring angles of filaments leaving spheres with 1, 1.5 and 2$\times R_{200}$. 

We also investigated whether filament alignments are influenced by the second most massive halo (SMH) in each simulation box -- as an indication for a possible mass transmission between them. In the simulations, the SMH  has halo masses of $M_{200} > 2.4 \times 10^{13}\Msun h^{-1}$, which is typically between 5\% and 30\% of the mass of the most massive halo and together, they represents a cluster pair (with a typical distance between the clusters of $9.7 \pm 3.5\hMpc$).  
Fig. \ref{fig:secondhalo} indicates that alignments of filaments are strongly influenced by the second most massive halo. These prominent \textit{bridges} between cluster pairs have historically been one of the first detections of filaments, marking especially strong and thick intra-cluster connections between close cluster pairs.
Such cluster-cluster bridges are believed to be remnants of large-scale filaments and with temperatures $T > 10^5 - 10^7 \rm{K}$, the gas emission of the hot ionised baryons have been detected in X-ray \citep{Vazza2019} as well as through the thermal Sunyaev–Zel’dovich effect \citep{Tanimura2019, Graaff2019}.

Filaments connect to nodes in a complex, multiscale manner. \citet{Ford2019} have shown that cosmic connectivity, i.e., the number of of filaments connected to a node (cluster or group) scales with the mass of groups and their brightest galaxies. High connectivity groups tend to have recently merged, which leads to a potentially interesting question of the dependence of connectivity with merger history or dynamical status. We intend to explore this question in the future. 

\begin{figure}
   \centering
   \includegraphics[width=\columnwidth]{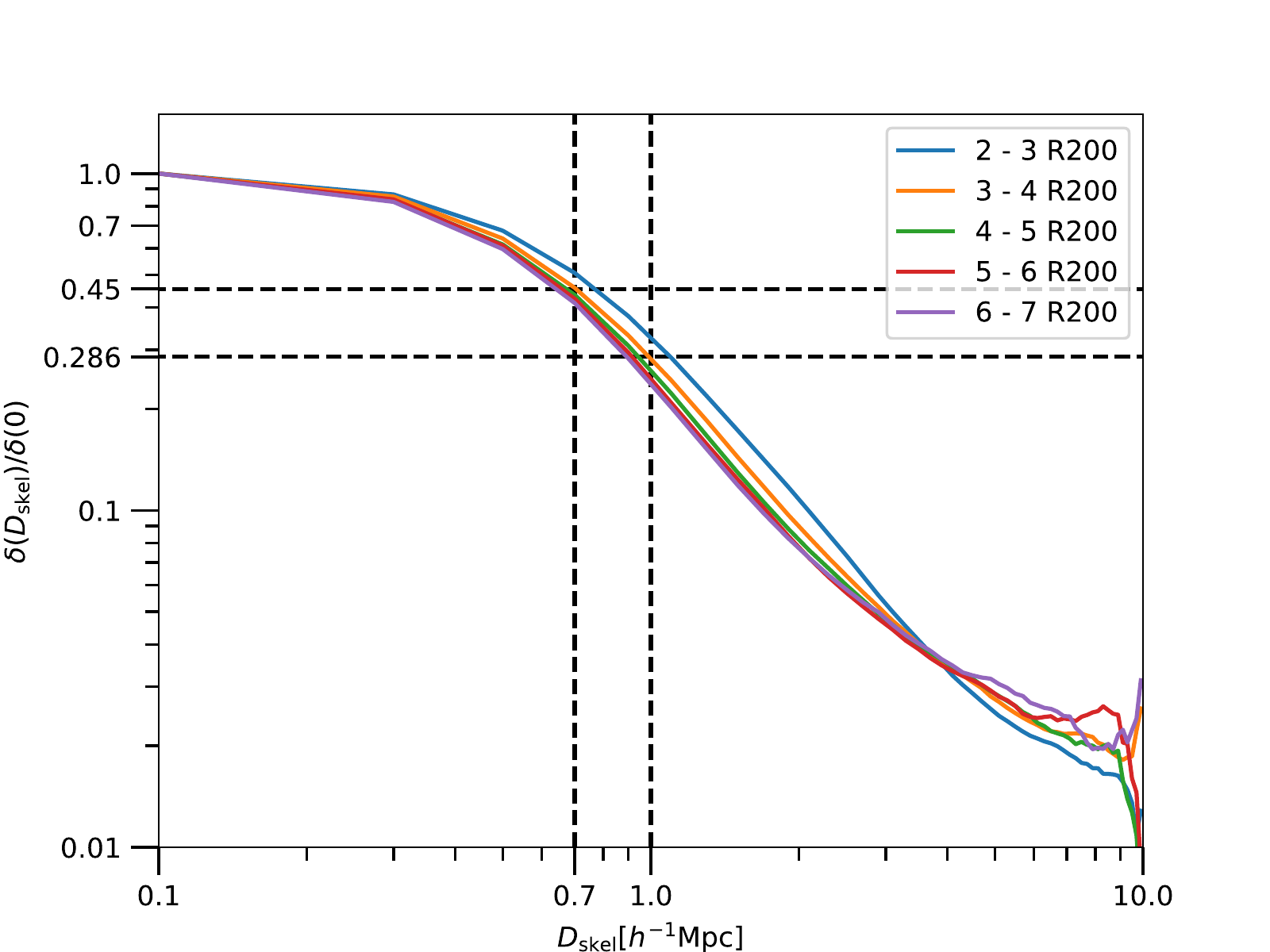}
    \caption{We define a characteristic thickness of filaments based on gas densities. Shown are the radial gas density profiles of gas filaments as a function of the distance to the filament centre ($D_{\rm{skel}}$). Different colours refer to distances to the cluster centre in steps of $R_{200}$. We exclude particles within $2R_{200}$ of halos in these filaments. In our work, we define filaments as cylinders with radius of $0.7\hMpc$. We compare results with a more relaxed radius of $1\hMpc$. Both are highlighted in the figure with dashed lines. The profiles are normalised by the density at the first bin ($0.1\hMpc$).
    }
   \label{fig:thickness}
\end{figure}

\subsubsection{Thickness of filaments}
\label{subsec:thickness}

While the cosmic web does have some thick, bridge-like structures (Sec. \ref{subsec:alignment}), it is dominated by small-scale filaments close to overdense regions, making the surroundings of clusters rich in thin filaments \citep{Cautun2014}.
The population of galaxies varies strongly with filament thickness \citep{Cautun2012}, in the sense that the thinnest filaments are mostly populated with low mass galaxies (due to the assembly bias) -- consequently making them harder to detect.
The thickness or boundary of filaments, defined by their radius or diameter, is therefore an important parameter to consider for galaxy evolution studies. 

In order to associate galaxies to filaments in our simulations that can be used in an observational setup as well, we first find a characteristic thickness of filaments around clusters. Note that this does not fully account for the multi-scale nature of the cosmic web and is a simplistic approximation within the likely limitations imposed by observations. 
We do this by defining the average filament radial density profile. The detailed procedure is described in Rost et al (in prep) and we advice the reader to refer to this publication for more information. In summary, they calculate overdensity profiles for the same suit of simulations for gas particles and dark matter particles. To deal with contamination of more massive halos, particles within $2R_{200}$ of halos were removed. This leads to an improved density contrast and allows to observe the pure underlying filamentary structure. 
Overdensity profiles of particles {\it p} were then determined as:

\begin{equation}
F_{\rm{p}}(r) dr = \frac{ N_{\rm{p}}(r, r + dr)} {N_{\rm{random}}(r, r + dr) }   \frac{ N_{0} m_{\rm{p}}}{V_{0}  \rho_{\rm{crit}}},
\end{equation}

where $N_{\rm{p/random}}(a, b)$ is the number count of {\it p}/random particles with perpendicular distance to the closest filament between $a$ and $b$, $N_{0}$ is the total number of random particles in the spherical region of the cluster, and $V_0$ is the total volume of that region.

In our work, we define filaments as curved cylinders with a fixed radius. Throughout the paper, we will compare results for filaments with radius $0.7\hMpc$ and $1\hMpc$.  
Unless otherwise stated, results and figures in this paper use a radius of $0.7\hMpc$ (section \ref{subsec:recoveryCM} explains this preference).
Other works have used a similar range of filament thicknesses \citep[e.g.,][]{Colberg2005, Tempel2014, Martinez2015, Sarron2019, Kooistra2019}. 
To quantify the effect of the different values, we investigate how much the density has typically dropped by a radius of $0.7\hMpc$ and $1\hMpc$, as seen in Fig. \ref{fig:thickness}. 
From the centre of the filament to $0.7\hMpc$, the gas particle density drops by an average factor of 2.2; i.e., the difference between the density in the filament centre ($\delta(D_{\rm{skel}})/\delta(0) = 1$), and the density at $0.7\hMpc$ distance from the centre ($\delta(D_{\rm{skel}})/\delta(0) = 0.45$). From the centre to $1\hMpc$, the density drops by a factor of 3.5 (dashed lines in the figure). These numbers change slightly depending on the distance to the node (i.e., distance from the cluster centre), as indicated by the coloured profiles that show bins along the filament length in steps of $R_{200}$. This means that the thickness of filaments in \threehundred simulations varies along the length of the filament with them being thicker closer to nodes. For example, a filament thickness with radius $0.7\hMpc$, the gas density has dropped by a fraction of 1.9 close to central node and a fraction of 2.4 furthest away from the node. 

Importantly, the \textit{shape} of the transverse profiles is very similar: whether we define the thickness close to the node, close to the saddle point or in between them makes only small differences that will be hard to distinguish in observations. Therefore, we use one average thickness along the entirety of the filaments, a more realistic assumption for our intentions. 
In addition, the density -- and therefore the derived thickness -- is similar between profiles measured on the basis of gas particles and of dark matter, where dark matter filament profiles are marginally thicker and more constant along the length of the filaments, i.e., the density varies less with the distance to the node (see Rost et al. in prep for a discussion of dark matter filaments in \threehundred.) 
\begin{figure}
   \centering
   \includegraphics[width=\columnwidth]{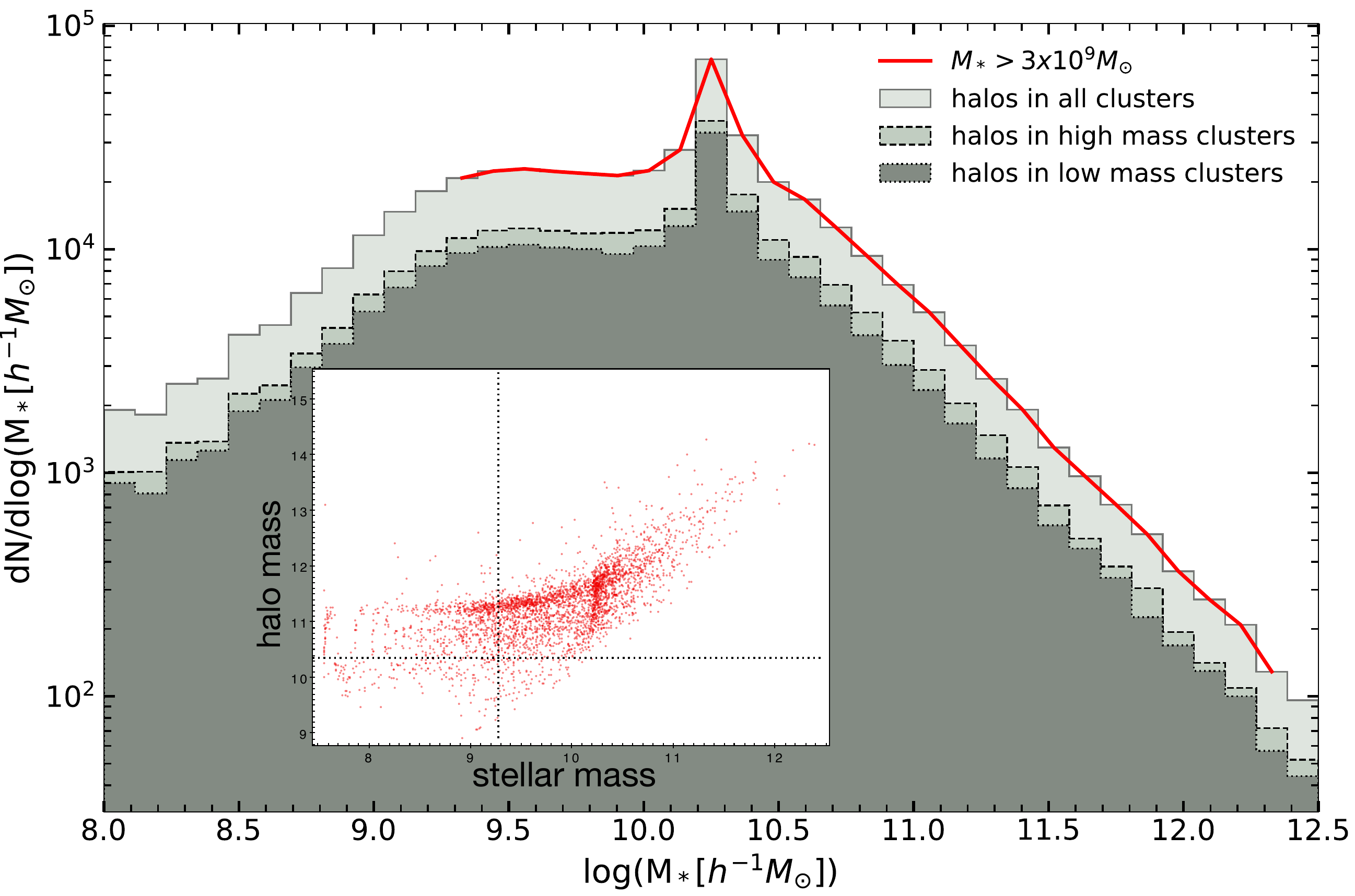}
    \caption {Referencing figure 7 in \citet{Cui2018}, the histogram shows the stellar mass function of halos selected to represent the future spectroscopic WEAVE Wide-Field Cluster Survey. We choose all simulated halos with $M_{\rm {halo}} > 3 \times 10^{10} \Msun$ (corresponding to $M_{*} > 3 \times 10^9 \Msun$ in this plot) inside a volume with length $15\hMpc$. The red line represents the mass distribution of halos in this mass range (mock galaxies) of all clusters. The medium shaded and dark shades show histograms for high and low mass clusters respectively, indicating that the distribution is independent of the cluster mass. The peak at $\rm{log}(M_*/h^{-1} Msun) \sim 10.3$ is caused by a combination of the simulation resolution and the striped/heated gas due to the Wendland kernel and AGN feedback in the \gadgetx simulations \citep[see][for details]{Cui2018}. The insert shows stellar masses plotted against halo masses of one cluster.} 
   \label{fig:mass-histo}
\end{figure}
Choosing $0.7\hMpc$ or $1\hMpc$ does not make a difference to our method. However, it is fair to point out that the thickness cut influences the results: by lowering the contrasts, filaments become thicker, the volume they occupy greater and consequently more galaxies are associated with them. Given the uncertainties of measuring filaments in an observational framework, our tests aim to find the \textit{optimal} thickness that provides a successful implementation to observations. 

\section{Towards Observations}
\label{sec:observations}

In order to assess the reliability and robustness of our filament extraction strategy for future surveys, we move from the idealised case of gas particles to mock galaxies, i.e., simulated halos that mimic galaxies with mass cuts comparable to those achievable observationally.
With MOS observations, spectroscopic redshifts can be used to allocate galaxies to structures -- a process that will allow to define volumes in observed space that are akin to the simulation boxes of \threehundred project. We therefore use halo catalogues from \threehundred simulations to reproduce conditions of spectroscopic surveys and compare the filaments detected using mock galaxies to our reference network that we have established from the underlying gas particles. We want to stress that at a very fundamental level, we expect galaxy and gas filaments to be different, and referring to the gas filaments as our benchmark framework is merely based on our aim to provide for future observations.
We especially highlight 
conditions of the future WEAVE Wide-Field cluster survey (WWFCS) as an imminent example, but also provide predictions for samples with a mass limit of higher-mass $L^*$-galaxies. The methods tested in this paper are therefore relevant and can be applied to other upcoming surveys, such as the 4MOST cluster survey \citep{Finoguenov2019}.

WWFCS will study 16 -- 20 cluster structures out to five $R_{200}$ in the redshift range $0.04 \leq z \leq 0.07$, with each 4000 -- 6000 galaxies within $5R_{200}$. 
WWFCS will thus cover the infall region with an unprecedented number of galaxies to date. This will be achieved through a mosaic of up to 20 pointings (with an average of 10 pointings, depending on cluster mass) of the 1000-fibre multi-object spectrograph WEAVE, that offers a field-of-view of 2\textdegree\, in diameter. The natural and most efficient target density is $\sim$900 targets per WEAVE field in the outer regions, which corresponds to $r=19.8$, and a stellar mass limit of $\sim$10$^9$ \Msun.
Here, we aim to test halos from \threehundred simulation similar to these observing conditions, both in mass range and in numbers. Taking both into account, we define mock galaxies with a minimum stellar mass of $M_{*} > 3 \times 10^9  h^{-1}\Msun$. In the simulation setup, this corresponds to halos with 
$M_{\rm {halo}} > 3 \times 10^{10} h^{-1}M_{\sun}$ inside a volume of radius $15 h^{-1} \rm{Mpc}^3$. We will refer to halos selected with these conditions as mock galaxies. This is illustrated in Figure \ref{fig:mass-histo} that shows the stellar mass function of halos from \threehundred clusters. Depending on the cluster, this yields between 2073 and 6636 simulated mock galaxies within $5R_{200}$, comparable to the number density expected for WWFCS volumes. 
In total, we find \mbox{$\sim 10^6$} mock galaxies outside 1 and inside $5R_{200}$ in the 324 simulation volumes combined. 

Note that WWFCS observations will provide spectroscopic redshifts instead of positions, which adds peculiar velocity-related distance errors affecting distance measurements. This "Finger of God" effect impacts filament finding, in particular close to the centre of clusters and is alleviated further away in cluster outskirts. This added uncertainty is not part of the current paper, and the topic of a future paper that will tailor specifically to observations of the WWFCS.

\subsection{Filament extraction using halos}
\label{subsec:fil_WEAVE}

\subsubsection{Mass-weighted mock galaxies filament extraction}
\label{subsec:mass-weighted}

For the 3D \disperse\, runs on the mock galaxy sample, we set a 5.3$\sigma$ persistence value and smooth the filaments with a smoothing parameter of 6 using x,y,z positions. While this extracts the majority of the filamentary network, in some cases central peaks (nodes) extracted in gas networks are not identified in mock galaxy networks 
However, for our analysis, nodes are important to quantify networks connected to the brightest cluster galaxies.
The discrepancy can easily be explained by the different natures of the input data sets: each gas particle is uniformly massive, however the gas particle number and distribution reflects a topological density field with peaks in high mass regions. For example, near the centre of each cluster, where we expect a massive brightest cluster galaxy (BCG) to dominate the field, many more gas particles are gathered than in regions of lower (gas-) density. The gas particle data set therefore effectively achieved a mass-weighting that defined nodes in areas of high number density, i.e., in high-mass regions -- something a realistic galaxy or halo point distribution cannot. However, this additional information is indeed present in observations where the brightness (luminosity) e.g., of the central galaxy gives additional valuable indication of the cluster topology. 

\begin{figure}
    \centering
    \includegraphics[width=\columnwidth]{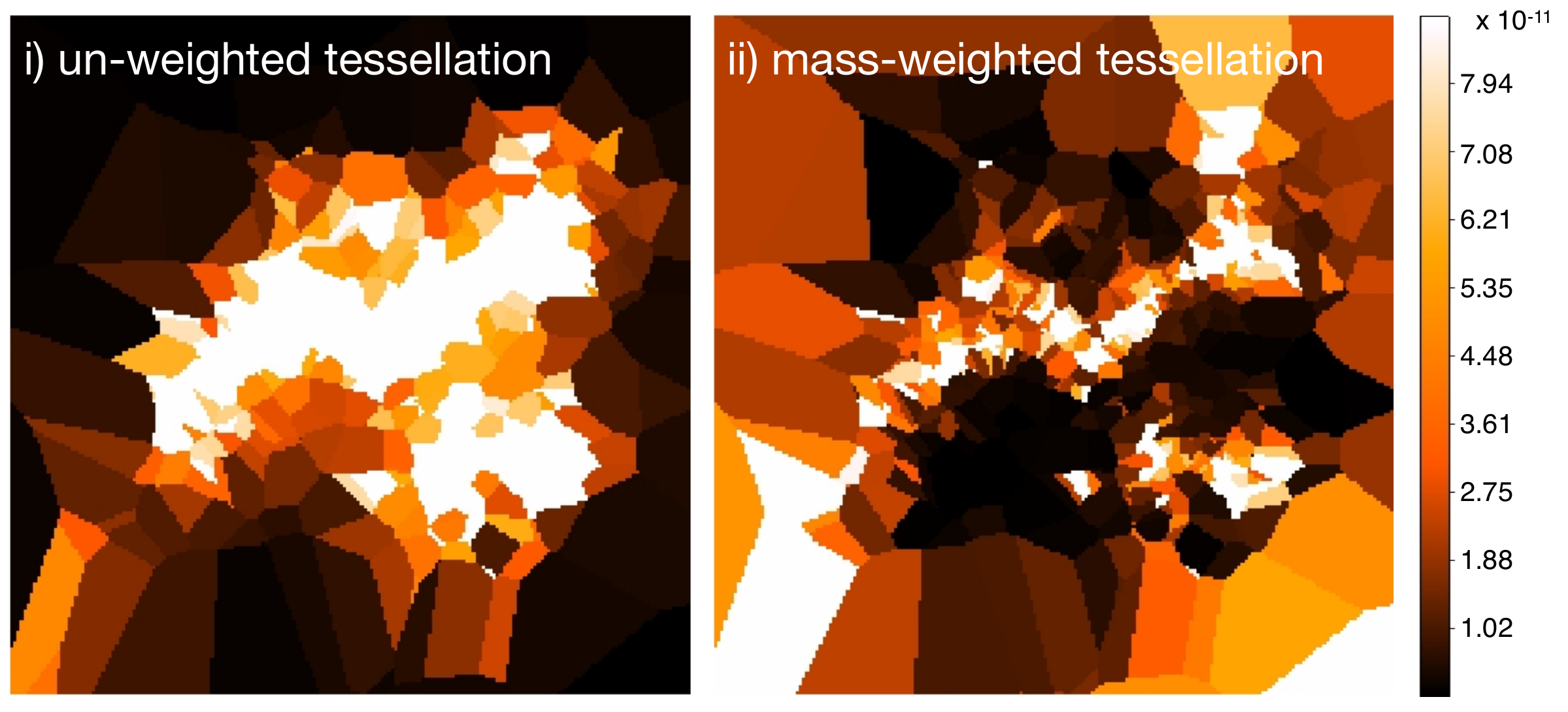}
    \caption{The figure highlights the impact the mass-weighting of halos has on the extraction of filaments. It shows the Delaunay tessellation, used by \disperse\ to identify filaments, in a slice of thickness 75~kpc around the centre of one cluster. Images are equally scaled. Units of the colour bar are arbitrary, but help to compare the two panels. Left: unweighted tessellation: all halos are equally weighted. Right: mass-weighted tessellation: the halos are weighted by their mass. We do this to achieve a closer resemblance to the gas distribution, our reference in this experiment (see Sec. \ref{subsec:fil_WEAVE} for details).}
    \label{fig:weighted_tessellation}
\end{figure}

In order to bring the skeleton extracted from mock galaxies in agreement with the gas extractions, 
we run \disperse\,again on a mass-weighted tessellation (see section  \ref{subsec:disperse} for explanation of how the Delaunay tessellation is employed). 
\footnote{For this, we compute the skeleton from a weighted tessellation, by tagging the tessellation using \disperse's "netconv -addField" option.}
This associates to each vertex of the tessellation a weight corresponding to the mass of the halo at this vertex. To be sure that the initial halos were well matched with the vertices, we matched their positions. This requires an adaptation of the persistence threshold, which we increase to 6.5$\sigma$.
Figure \ref{fig:weighted_tessellation} shows the impact this mass-weighting had on finding filaments. It is a visualisation of the tessellation onto a cartesian grid of a slice of 75kpc thickness  
around the centre of a simulation box. The left panel shows the tessellation without mass-weighting and the right panel clearly reveals how the mass-weighting helped with the identification of filaments.
With this additional step, we accomplished our goal to identify all central nodes (BCG's), which we used to specify the main networks of each cluster. Note that in an observational setup, the weighting can be achieved in similar ways using observed luminosities or estimated stellar masses. 

Finally, we repeat the feature extraction using the projected density field of mass-limited halos, providing 2D coordinates as inputs to \disperse, and adjusting the persistence threshold and skeleton smoothing parameters to 3.2 $\sigma$ and 60 respectively.

\subsubsection{$L^*$-galaxies filament extraction}
\label{subsubsec:Lstar}

The best tracers of filaments are massive galaxies. Studies have shown that galaxies are more massive closer to filaments than further away \citep{Malavasi2016, Kraljic2017, Chen2016, Sarron2019, Bonjean2019}. 
We therefore also explore the possibility of using a higher mass limit as accessible tracers of filaments. 
However, at higher masses, the number of objects decreases rapidly. Following suggestions in \citet{Robotham2013}, we therefore define our \mbox{$L^*$-galaxies} sample as all mock galaxies with stellar masses greater than $10^{10}\Msun$. This conservative mass limit also comfortably includes galaxies with stellar masses similar to the Milky Way galaxy (MWG) with $5 \times 10^{10} \Msun$ \citep{Flynn2006}.

While this mass (or luminosity) threshold offers a high contrast and is available for most surveys, the trade-off is that the density is less well sampled.
By construction, this only includes high mass galaxies and therefore reduces the number to between 400 and 1100 objects per cluster. Note that this number is already available for several existing cluster surveys (e.g., CLASH \citep{postman12}; LoCuSS \citep{Haines2013}; Hubble Frontier Fields \citep{Lotz2017}; Omega-WINGS \citep{Moretti2017}). We adopted \disperse\ parameters to a persistence $\sigma = 4$ and a skeleton smoothing parameter of 5.

The parameter values used for all mock galaxy Disperse runs were identified by minimising the value through the extraction assessment described in the following section. In practice that means that we repeated the assessment multiple times, each time updating the values based on the previous result. The best value finds filaments and critical points similar to the reference framework. 


\subsection{Extraction Assessment}
\label{subsec:assessment}
\begin{figure}
   \centering
   \includegraphics[width=\columnwidth]{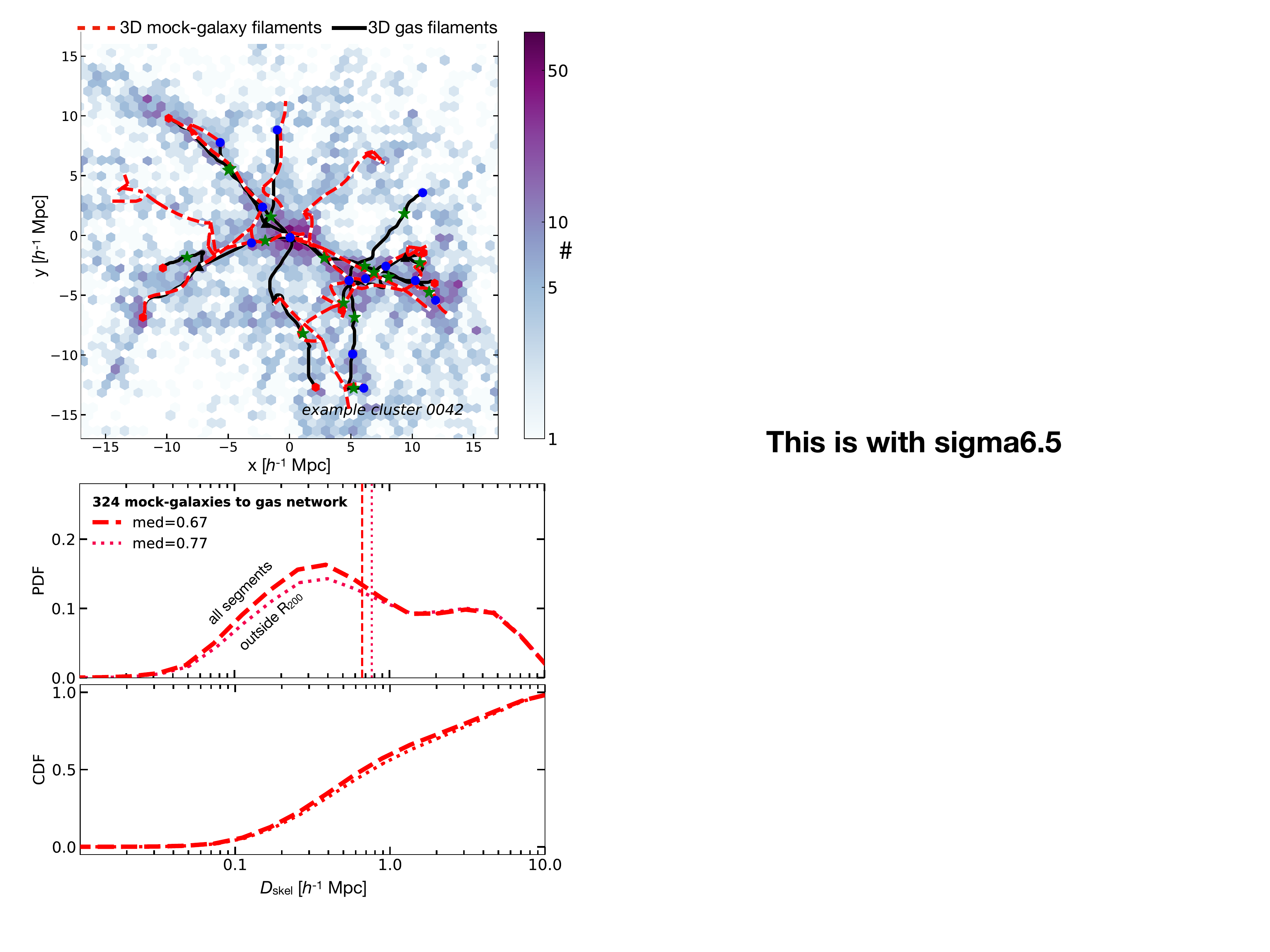}
    \caption{Top: Comparison of extracted filaments from the mass-weighted mock galaxy distribution in 3D to the underlying 3D gas-particle distribution (our reference skeleton) of one example cluster. Filaments extracted from 3D mock galaxies are plotted in red dashed lines, those extracted from smoothed gas particles are black. Nodes, saddle points etc. are marked as described in Figure \ref{fig:stability}. As can be seen in this example, some filaments do not have a counterpart (see text for discussion). Probability (middle panel) and cumulative (bottom panel) distribution of the distances between skeletons of filament networks from mock galaxies and gas particles for the entire sample of 324 clusters. Dotted lines use segments outside $R_{200}$, dashed lines include them. Vertical lines show medians of the distances between filament extractions, values are printed in the legend.}
   \label{fig:gas_vs_weightedWEAVE}
\end{figure}
\begin{figure}
   \centering
   \includegraphics[width=\columnwidth]{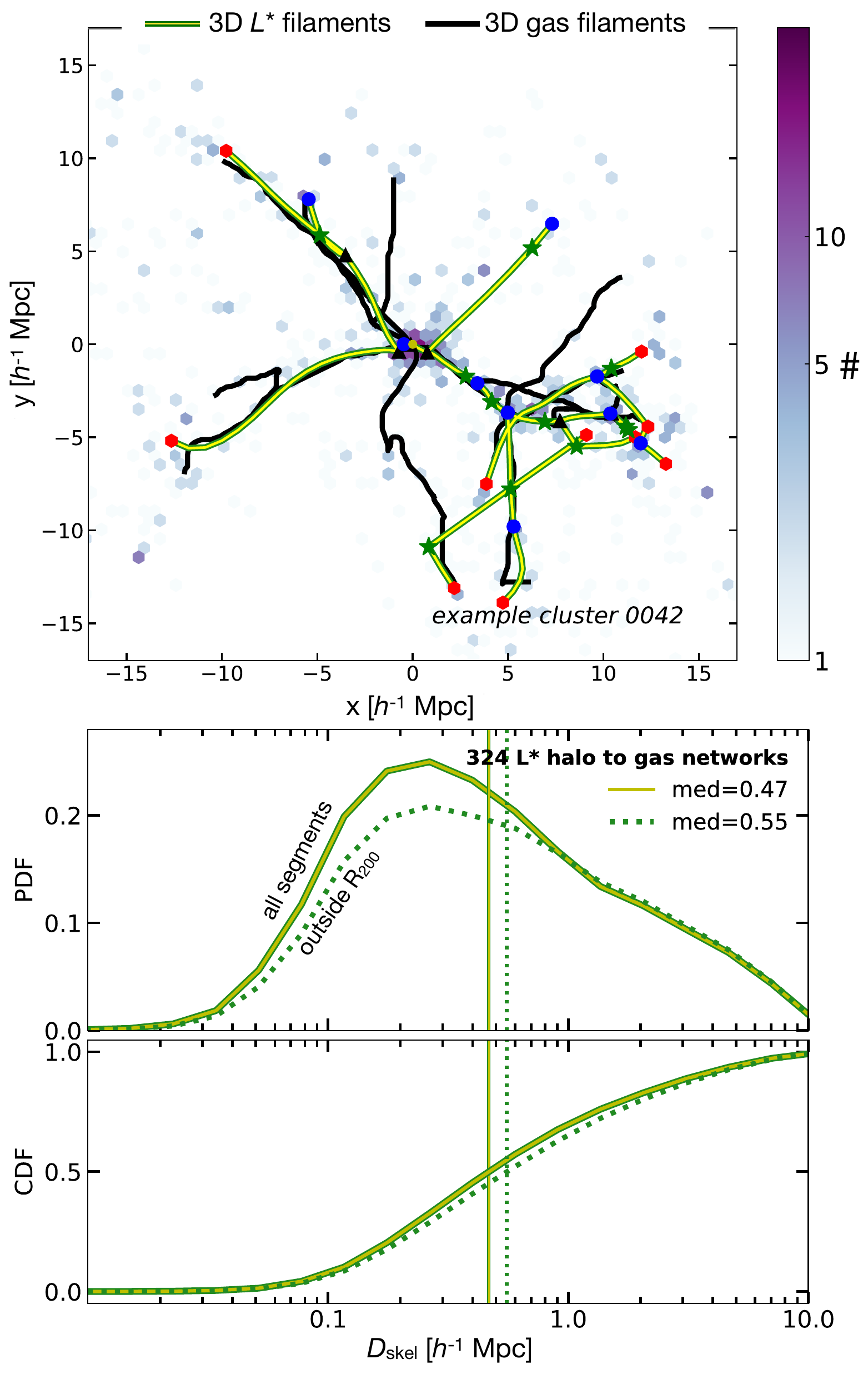}
    \caption{Top: Comparison of filaments extracted using a higher mass limit of $M_* > 10^{10}\Msun$, equivalent to $L^*$-galaxies, with the reference extraction based on smoothed gas particles. We use the same cluster as in Figure \ref{fig:gas_vs_weightedWEAVE} for this example. Points as explained in \ref{fig:stability}. Middle and bottom panels: PDF and CDF of the distances between skeletons of filament networks from $L^*$- and gas network of all clusters combined. Solid lines use segments outside $R_{200}$, dotted lines include them. Vertical lines are medians, and values are printed in the legend.}
   \label{fig:Lstar_gas}
\end{figure}

\subsubsection{Comparison of 3D gas filaments to 3D mock galaxy filaments}
\label{subsec:comparison_gas_halos}

First, we compare skeletons extracted from the 3-dimensional distributions of gas particles (our reference network) to 3-dimensional (mass-weighted) mock galaxies. This is illustrated in Fig. \ref{fig:gas_vs_weightedWEAVE}. The top panel shows a projection of these two filament networks for one typical example cluster. Filaments extracted from gas particles are shown in black solid lines and filaments extracted from the mass-weighted mock galaxy distribution are shown in red dashed lines. They are plotted on top of the (projected) mock galaxy distribution, shown in colour-coded hexagonal 2D-histograms. 
It is no surprise that, typically, they do not match perfectly, because 1) the mock galaxy distribution is already a biased tracer of the underlying density field and 2) we have far more gas particles than halos leading to a more precise density field, which in turn leads to a more accurate filament extraction. As explained in Sec. \ref{subsec:fil_WEAVE}, we try to counteract this by weighting by mass.
Despite their very different inputs, the two are in relative good agreement throughout our sample of 324 cluster simulations. The example chosen for Fig. \ref{fig:gas_vs_weightedWEAVE}, however, also clearly shows that some filaments do not have counterparts in the respective other skeleton at all: they are recovered in one, but not the other density field. These spurious detections directly result from the choice of parameters - a trade off that is difficult to bypass, as noted in \citet{Laigle2017}. 

In the bottom panel of Fig. \ref{fig:gas_vs_weightedWEAVE}, we quantify the discrepancy/similarities between the 3D-gas- and 3D-halo skeletons \textit{over the whole ensemble of clusters}. We follow a method that was introduced in \citet{Sousbie2011} and used in \citet{Laigle2017} and \citet{Sarron2019} and offers an indication of the reliability of the filament extraction.
For this, we measure the distances between the two skeletons in all cluster simulations and plot their differential distributions (PDF) and cumulative distribution (CDF). In this section, we compute the distances in 3D between each segment in the mock galaxy network and the nearest segment in the gas network in each of the 324 clusters.
The dashed line shows the resulting PDF and CDF of distances of the sum of all skeletons (i.e., using all segments for 324 clusters) and the dotted line is the result for all skeletons outside $R_{200}$. The corresponding vertical dashed lines give the medians of the two distributions: $0.67\hMpc$ and $0.77\hMpc$ for all segments and for segments outside $R_{200}$ respectively. Medians are always higher when excluding the contribution of segments inside $R<R_{200}$. This is because inside $R_{200}$, segments lie close to each other because the volume is small. 
It is encouraging that these numbers are comparable to previous measurements from larger simulations found in the literature \citep{Laigle2017,Sarron2019} and that the majority of the distances are lower than the typical thickness of a filament. Note, however, the long tail and even extra bump in the distributions. This shows that there are filaments that do not have a counterpart at all.
\bigskip

We also want to test whether using a more accessible higher mass limit for galaxies can recover the filament network. Evidence shows that high mass galaxies are found closer to filaments, suggesting that they could lead to a more robust extraction, even in cases where lower mass galaxies are available.
Despite the drastic reduction in numbers compared to mock galaxies fed to \disperse, we found a good agreement of filaments from $L^*$ galaxies ($M_* > 10^{10}\Msun$) to the filaments extracted using gas particles (Fig. \ref{fig:Lstar_gas}). As a reminder, the $L^*$-sample uses 400--1000 mock galaxies for filament extraction, the weighted mock galaxy sample with lower mass limits of $3 \times 10^9 \Msun$ uses 3000--6000 objects. 
Our experiment shows that using $L^*$-galaxies as tracers robustly recovers the main filaments of each network. This works especially well when the system is simple. However, in some clusters (less than 10\% of our sample), the main node was not identified -- just as we found when using a lower mass limit without weights.
If this is necessary for the analysis of the science case, we suggest a mass- or luminosity-weighted approach as outlined above. 

If the main goal is purely to find the main filament network, then using a sample of high mass galaxies with a (conservative) $L^*$-mass limit of $M_{*}>10^{10}\Msun$ is a good approach that achieves comparable results for finding filaments, while being accessible and straight forward to use.
We show this quantitatively in the lower panel of Fig. \ref{fig:Lstar_gas}. As in Fig. \ref{fig:gas_vs_weightedWEAVE}, this is the PDF of distances between segments. The median of these distances is $0.47\hMpc$ for all segments and $0.55\hMpc$ for segments outside of $R_{200}$. Our assessment shows that, given our choices, the median distances between the reference network (i.e., gas filaments) and the $L^*$-galaxies filaments is smaller than for halos with lower mass-limits. 
\begin{figure}
   \centering
   \includegraphics[width=\columnwidth]{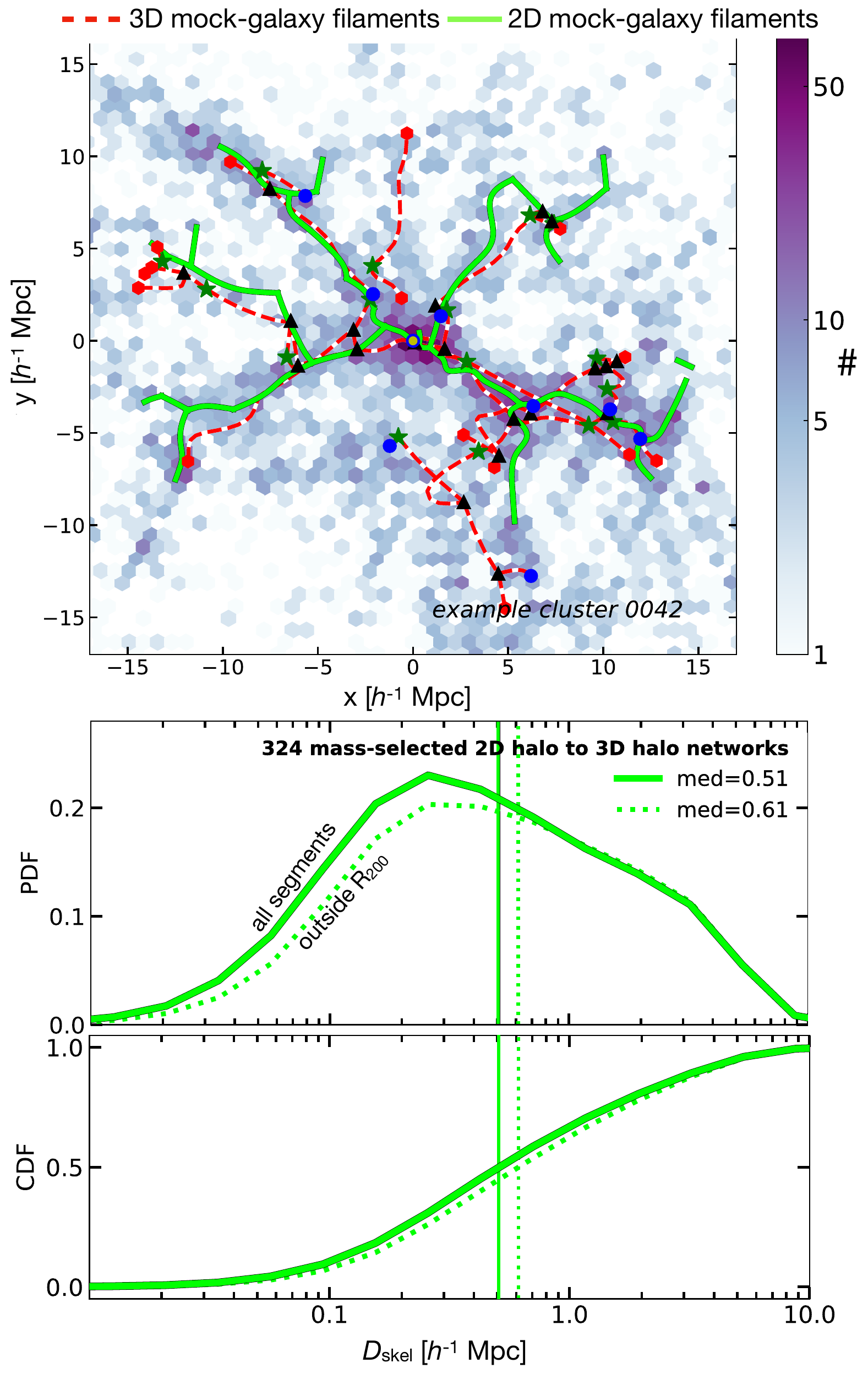}
    \caption{Top: Comparison of filaments extracted using the mass-weighted 3D mock galaxy distribution with the 2D-projection of the same halo sample for one cluster example -- the same cluster as in Figure \ref{fig:gas_vs_weightedWEAVE} and \ref{fig:Lstar_gas}. Points as explained in \ref{fig:stability}. Bottom: PDF and CDF of the distances between skeletons of filament networks from 3D and 2D of all clusters combined. Solid lines use segments outside $R_{200}$, dotted lines include them. Vertical lines are medians, and values are printed in the legend.}
   \label{fig:3D_vs_2D}
\end{figure}
\begin{figure*}
    \centering
    \includegraphics[width=1\linewidth, keepaspectratio]{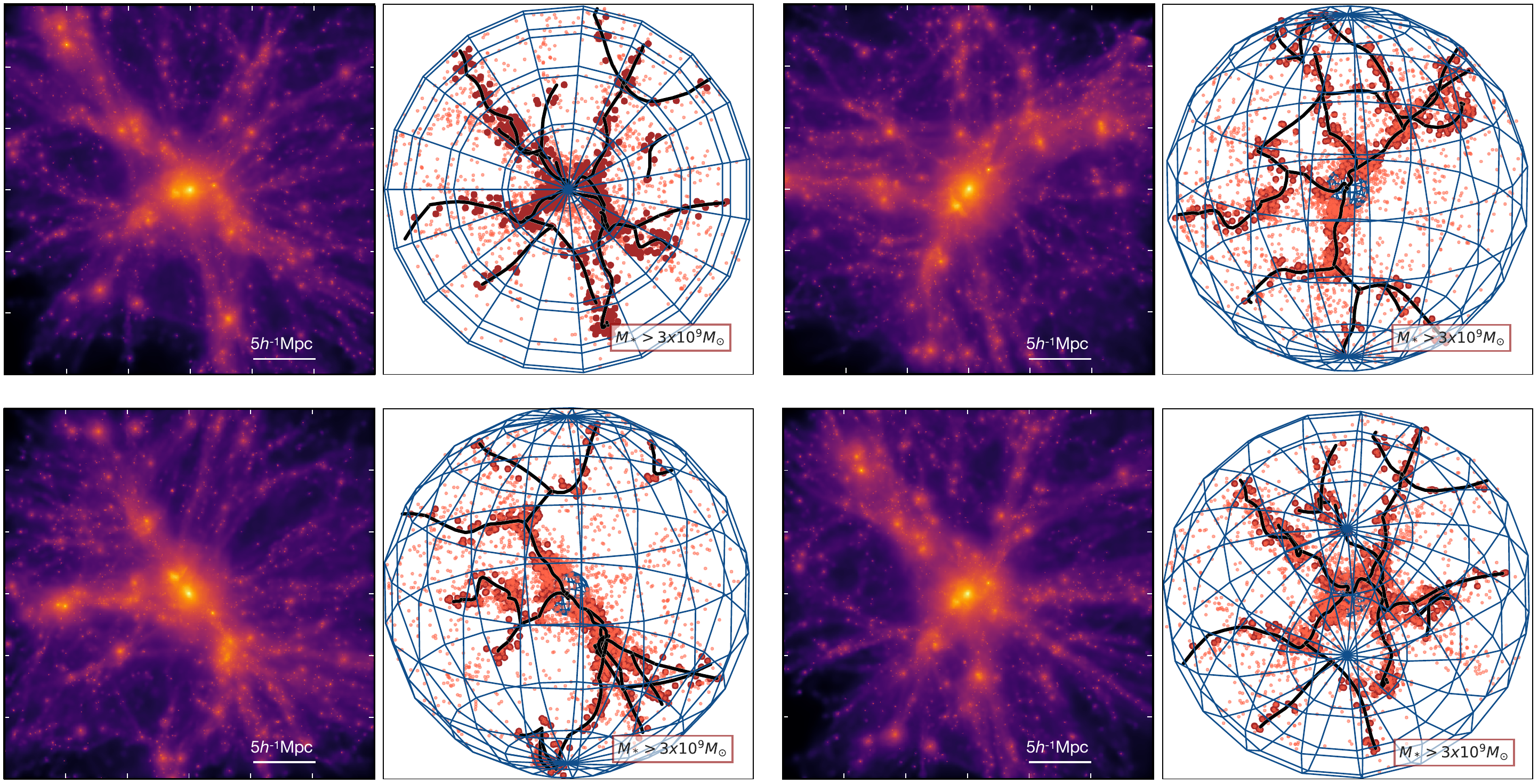}
    \caption{One (random) example cluster of \threehundred Project depicted at four different angles. Each pair shows the cluster in gas particles (left) and \disperse\ filament network with associated mock galaxies (right). The filament network was extracted from the distribution of mock galaxies with $M_*>3\times 10^{9} \Msun$.}
    \label{fig:fil_assoc}
\end{figure*}

Filaments are biased towards more massive galaxies. This is the reason why either a weighting by mass, or choosing higher mass galaxies will yield robust results. Furthermore, galaxies with a higher signal-to-noise ratio will be better tracers for the filament finding algorithm.
We conclude from our experiment, that choosing mock galaxies with $L^*$-mass limit offers an ideal contrast for \disperse\ to find the main filaments around clusters. However, note that only a weighting (in our case) by mass guarantees the correct definition of nodes in all clusters without human intervention.
The weighing offers a hands-off filament finding method that correctly identifies nodes, without making decisions a priori of selecting the brightest cluster galaxies in observations or tagging them in simulations. 


\subsubsection{Comparison of filaments extracted from mock galaxies in 3D and projected 2D}
\label{subsec:comparison_3D_2D}

In an effort to get one step closer to observations, in particular when reliable spectroscopic redshifts are not available, we now compare the skeleton extraction using the 3-dimensional distribution of mass-selected mock galaxies to skeletons extracted from the same distribution, but now projected onto the x-y-plane.
The top panel in Fig. \ref{fig:3D_vs_2D} compares filament networks of the same typical cluster as in Figures \ref{fig:gas_vs_weightedWEAVE}: red dashed lines once again show filaments extracted from mock galaxies in 3D, and green solid lines are the results of \disperse\ using the 2D mock galaxy distribution. 

Qualitatively, the two networks agree well. We repeat the same procedure as described in the previous section to assess the filament extraction statistically for all clusters combined. We establish the distribution of distances between the two cleaned networks by calculating the minimum projected distances between each segment of the 2D filament network with the 3D filament network, repeatedly for the entire cluster sample. The result for all clusters is shown in the lower panel of figure \ref{fig:3D_vs_2D}. Most 2D filaments are reliable counterparts of 3D filaments. As before, we take the median of the two distributions as a quantitative measure of the reliability of the filament extraction in 2D compared to 3D. 
We find that on average, the segments of the 2D filament network are $0.61\hMpc$ distant for filaments outside of $R_{200}$ and $0.51\hMpc$ including filaments inside $R_{200}$.
This second number is slightly larger than what was found in \citet{Laigle2017} and \citet{Sarron2019} in the case of large simulation boxes ($0.32\hMpc$ and $0.34 \hMpc$ respectively). However, their numbers are expected to be lower than ours, since a large simulation box leads to stronger projection effects and 2D filaments appear more closely together.
A more comparable approach is the one used in \citet{Sarron2019} that focuses specifically on filaments connected to clusters (and up to the first saddle point). In this case, they find a median distance of $0.55\hMpc$ (including filaments inside $R_{200}$). Note, however, that even though this is very similar to what we find, they are using slices in redshift space 20 times as deep as our volume, again increasing projections.


For this exercise, we used coordinates of halos once in 3D (x,y,z) and once in 2D (x,y) which require different $\sigma$-thresholds. This means that the input parameters vary, which can explain some of the differences. However, the far more obvious cause for differences are projection effects in 2D that are misinterpreted as peaks in the density distributions. In projection, filaments could connect points that may be spatially separated in 3D. 

Comparing the previous two sections, we can see that, at least for our sample, the step from millions of particles to thousands of halos impacts the reliability of filament extraction more than the projection from 3D to 2D. 

\subsection{Mock galaxies associated to filaments and their dependence on cluster properties}
\label{subsec:association}

By answering three key questions, the next three sections aim to fully link simulations to future observations. 
We want to know: (1) What is the fraction of galaxies in filaments in an idealised simulated (3D) environment and how does this change with simulated detection limits? (2) Does this number depend on cluster radius? (3) What changes in a realistic observational (2D projected) setup?  

Fig. \ref{fig:fil_assoc}  illustrates our path from simulated galaxy clusters to mock galaxies associated to filaments.  The left image of each pair shows our starting point: the gas particle distribution of one example cluster viewed from four different angles. The right panels show the halo distribution of the same cluster and at the same rotations. Small points show the positions of all mock galaxies with \mbox{$M_* > 3\times10^9 \Msun$} outside the cluster's $R_{200}$; highlighted are halos associated to filaments. The illustration shows filaments that we extracted using the weighted mock galaxy sample (in black).

\subsubsection{The impact of filament extractions}

In this section, we compare filament extractions and fractions of associated galaxies for a variety of observationally relevant setups.
Specifically, we assess filament extractions using (1) smoothed gas-particles as well as galaxies with mass-limits of (2)  $M_*>3\times10^9 \Msun$ (mock galaxies for short) and (3) $M_*>10^{10} \Msun$ ($L^*$-galaxies).We further investigate fractions of galaxies with the mass-limits corresponding to (1) the mock galaxies, (2) the $L^*$-galaxies, and (3) MW-like galaxies. We discuss results for filament radii $0.7\hMpc$ and $1\hMpc$. 

\begin{figure}
    \centering
    \includegraphics[width=0.98\columnwidth]{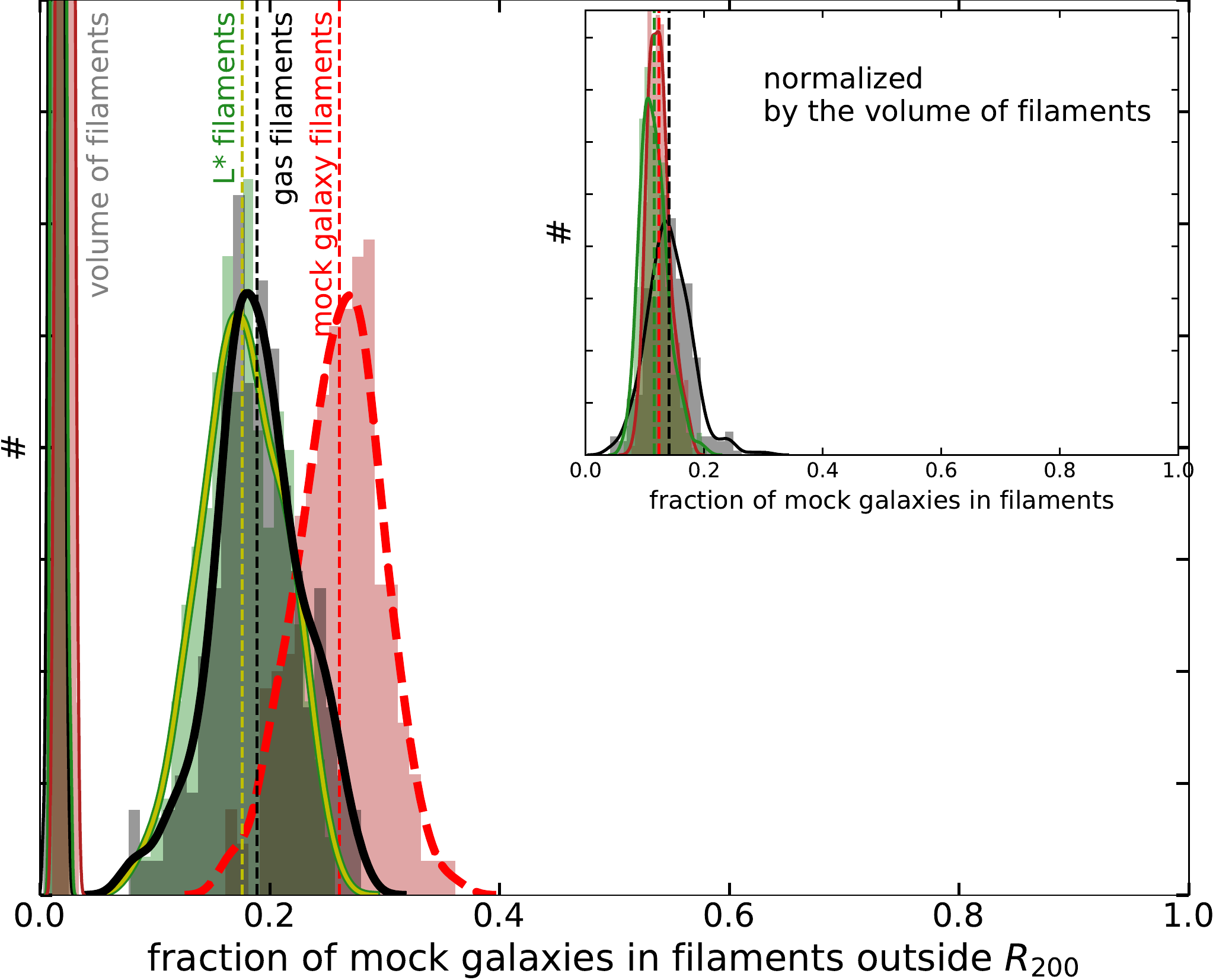}
    \caption{The fraction of mock galaxies (halos with $M_*>3 \times 10^9 \Msun$) in filaments ($D_{\rm{skel}}<0.7 \hMpc$) varies by $\sim10\%$ depending on different filament extractions. We show histograms of the fraction of mock galaxies in gas-filaments drawn for all 324 clusters in black, mock galaxy-filaments in red and $L^*$-filaments in green. Dashed lines are the mean values. Also shown is the fraction of the total volume that the filaments occupy. We use this to normalise the fraction of mock galaxies associated to filaments. This is shown in the insert. Outside $R_{200}$, filaments occupy between 2\% and 5\% of the volume cluster infall region, but contain up to a quarter of the mock galaxies. This reduces to $\sim15$\% for all extraction methods when we normalise the fractions by the volumes.} 
    \label{fig:fil_vol}
\end{figure}

Fig. \ref{fig:fil_vol} shows the fractions of mock galaxies in filaments for 324 clusters extracted from gas (black histogram), $L^*$-galaxies (green histogram) and mock galaxies (red histogram). The dashed lines are the mean values for each filament extraction method. 
On average, $\sim19$\% of mock galaxies are associated to gas filaments, $\sim17.5$\% are associated to $L^*$-defined filaments and $\sim26$\% are around mock galaxy-extracted filaments.   
The figure also shows the fraction of the total volume that the filaments occupy. Only a few percent of the volume outside $R_{200}$ (2\% for $D_{\rm{skel}}<0.7 \hMpc$ and 5\% for $D_{\rm{skel}}<1 \hMpc$) are occupied by filaments, but they contain up to a quarter of all mock galaxies.

The insert shows galaxy fractions for each filament-finding method normalised by the volume they occupy. 
Gas and $L^*$ filaments occupy similar volumes and trace a similar fraction of mock galaxies. 
Mock galaxy filaments have a higher fraction of galaxies ($\sim26$\%), but also occupy more volume. This is evident in the insert, where the red line jumps from the highest fractions to having fractions similar to the $L^*$ and gas networks. Evidently, our mock galaxy extraction is passing through regions with galaxies more frequently than gas or $L^*$ filaments. This means that -- despite our efforts to replicate filaments based on gas particles -- our filament finding based on mock galaxies does not carve out the same galaxy-filled regions as the filaments based on gas particles; it carves out more volume and finds more galaxies. Normalised by the volume, all filament finders find a similar fraction of mock galaxies: between $\sim12$\% ($L^*$ and mock galaxy filaments) and $\sim14$\% (gas filaments). While the difference is minimal, it shows that gas filaments are most successful in tracing regions dense in galaxies. 
While this leads to some contamination in the characterisation of the filament network, it adds very little contamination to the galaxies in filaments. We speculate that this discrepancy is due to the persistence threshold we chose. 

\subsubsection{The impact of detection limits and cluster properties}
\begin{figure}
  \centering
  \subfloat[dependence on cluster mass]{\includegraphics[width=0.49\textwidth]{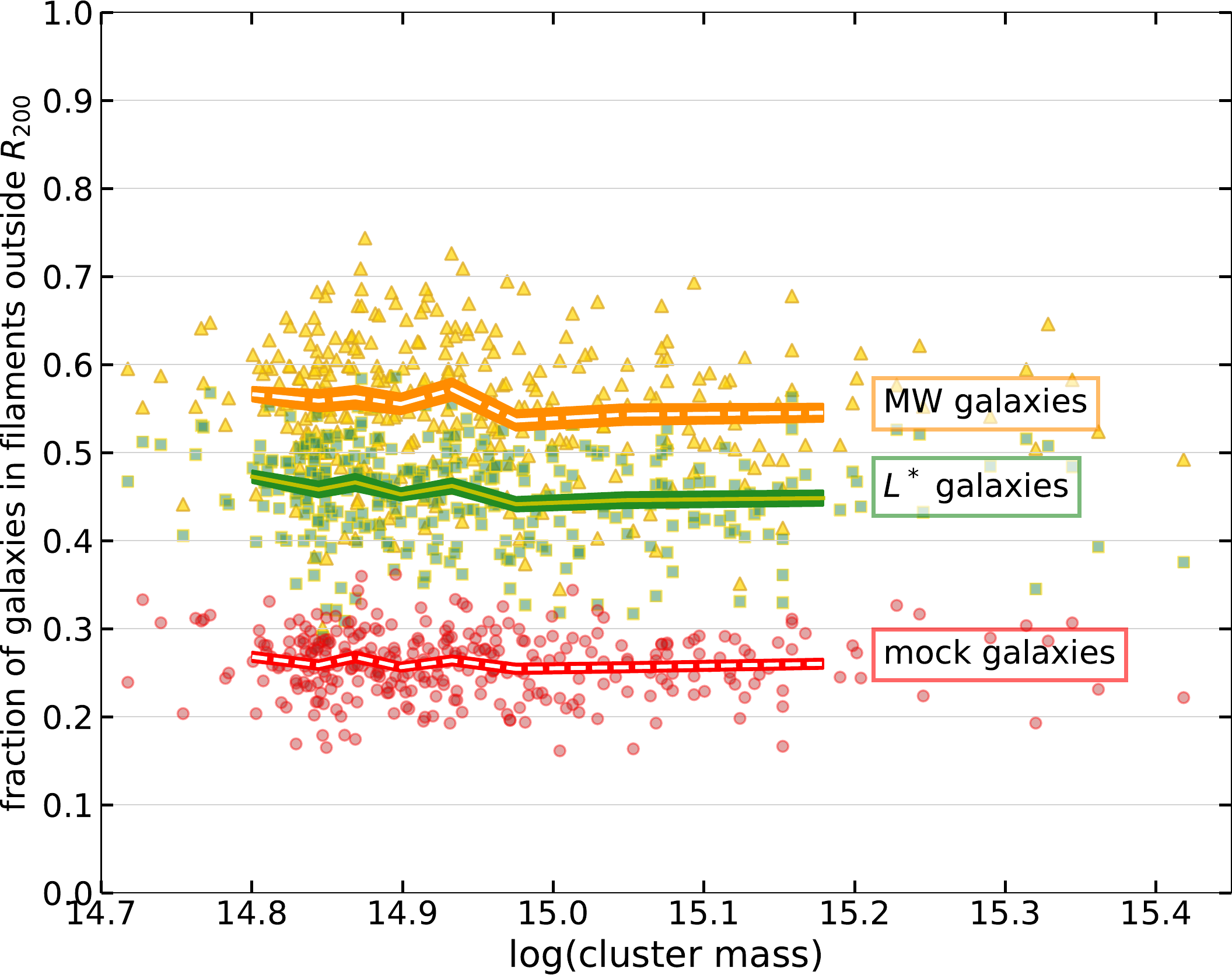}\label{fig:frac_mass}}
  \hfill
  \subfloat[dependence on dynamical state of the cluster]{\includegraphics[width=0.49\textwidth]{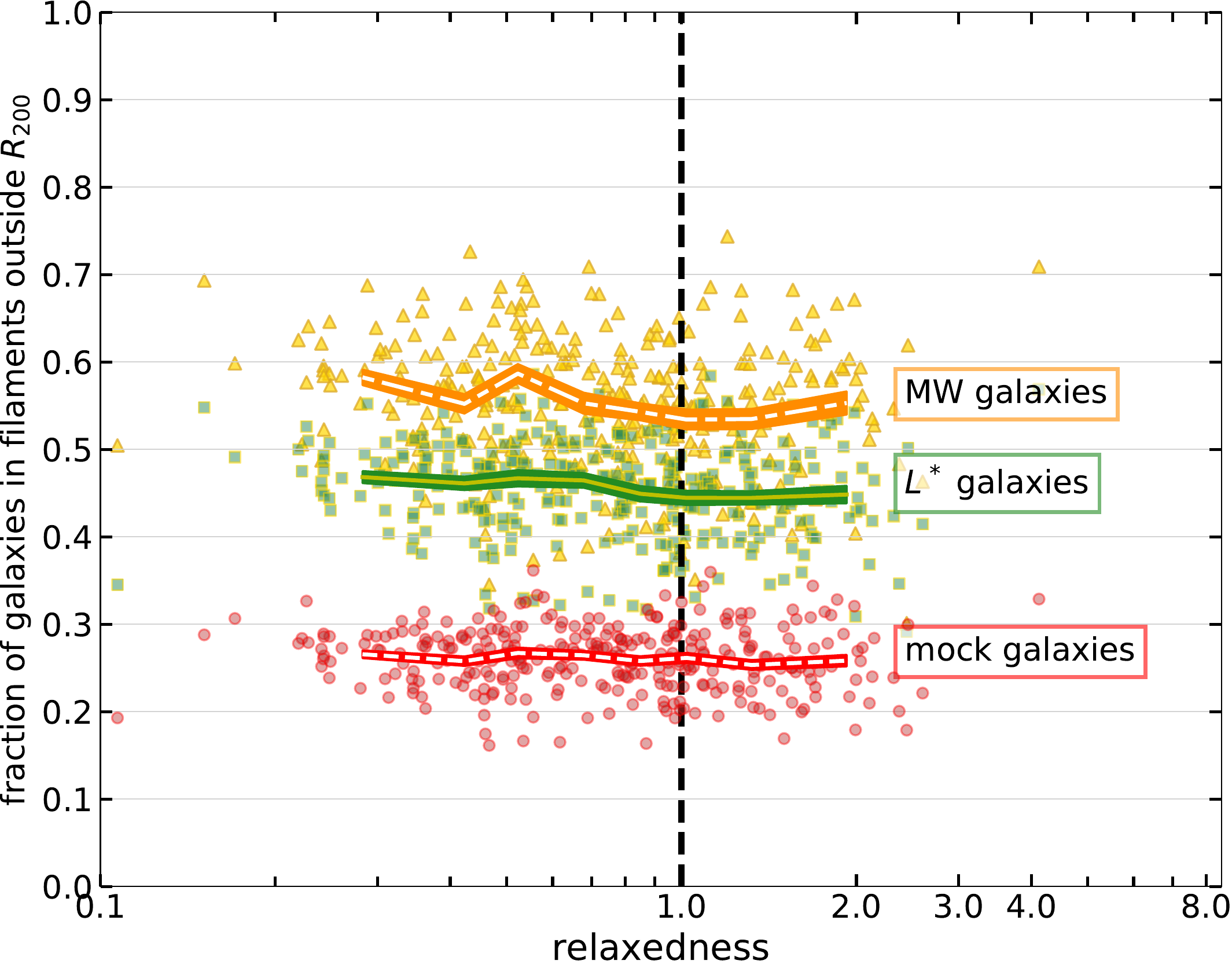}
  \label{fig:frac_relaxedness}}
  \caption{The fraction of halos in filaments does not depend on mass or dynamical state of the cluster. However, the fraction changes dramatically with galaxy mass. Shown are three mass cuts: Milky Way-type simulated galaxies with $M_*> 5\times 10^{10}$ in orange dot-dashed lines, $L^*$-galaxies with $M_*> 10^{10}$ in green solid lines and the lower mass-selection of mock galaxies with $M_*>3 \times 10^9 \Msun$ in red dashed lines. Coloured bands are 1$\sigma$ error on the mean. About a quarter of all mock galaxies are associated to filaments, whereas more than half of all Milky Way-type galaxies are found in filaments.  }
  \label{fig:fractions}
\end{figure}
Does the number of galaxies in filaments depend on the sample depth or cluster properties?   
Fig. \ref{fig:fractions} shows the fractions of galaxies in filaments ($D_{\rm{skel}}<0.7 \hMpc$) outside the cluster's $R_{200}$ as a function of cluster mass (Fig. \ref{fig:frac_mass}) and relaxedness (Fig. \ref{fig:frac_relaxedness}).  Each point represents the fraction of galaxies in weighted mock galaxy filaments of one cluster, while the bands indicate the means of the point distributions and corresponding errors.
The fraction of galaxies in filaments is galaxy-mass dependent. For a given filament extraction method, massive galaxies are more likely to be in filaments than outside filaments.
More than half of all Milky Way-type galaxies belong to filaments (55.8\%, orange dot-dashed line). 
This fraction drops to 46.3\% in $L^*$ galaxies and to $\sim$26.5\% in mock galaxies with $M_*>3\times10^9\Msun$ (red dashed line).
Naturally, the numbers increase if we increase the thickness of the filaments: the MW-galaxy fraction increases to 60.8\% and the mock galaxies fraction increases to 30.8\% for $D_{\rm{skel}}<1 \hMpc$.
This galaxy-mass dependence is a manifestation of the observed transverse stellar mass gradient of galaxies towards filaments, i.e. massive galaxies are closer to filament centres than less massive galaxies \citep{Malavasi2016, Laigle2017, Kraljic2017}. These studies have also shown that even on large cosmic-web scales and when the contributions of the nodes (clusters) are removed, mass gradients towards filaments prevail.

The figures further show that the fraction of galaxies in filaments does not depend on the mass (Fig. \ref{fig:frac_mass}) or on the dynamical status of the cluster, as expressed by the relaxedness parameter (Fig. \ref{fig:frac_relaxedness}). Cluster mass grows self-similarly. 
This is true for all filament extraction methods and galaxy mass limits that we tested. Note that the total \textit{number} of galaxies in filaments increases with cluster mass, but the \textit{fraction} stays the same. This is because the galaxy number density is higher around more massive clusters. At the same time, because massive clusters are usually more unrelaxed (Sec. \ref{fig:relaxedness}), the \textit{number} of galaxies in filaments decreases with relaxedness, but not the fraction. 

The dynamical state (relaxedness) is not intrinsic or fundamental to the cluster, but evolves over time. Processes in their recent history since $z=0.4$ are crucially effecting their composition at the present day. \citet{Haggar2020} have shown that unrelaxed, dynamically active clusters have been accreting a large amount of material in the last few Gyrs, which we might expect to increase the fraction of galaxies in the filaments around them. 
However, because the clusters rapidly grow their $R_{200}$, the population of galaxies in filaments close to $R_{200}$ is incorporated by the growth of the cluster. Consequently, we do not see a higher fraction of galaxies in filaments in unrelaxed clusters (Fig \ref{fig:frac_relaxedness}).

\subsection{A pile-up of galaxies in filaments closer to cluster centres}
\label{subsec:radial_filaments}

The previous analysis showed the mass-dependent fraction of galaxies associated with filaments using one average value for every cluster volume. In the following section we investigate whether the fraction of galaxies in filaments depends on the radial distance to the cluster centre. 
In addition to an increase of the galaxy density towards the cluster centres, we also expect galaxy mass gradients driven by the local mass-density relation, making more massive galaxies more prevalent in dense regions. Because in addition to these local effects, massive galaxies are also closer to filaments as a secondary driver, we may expect a higher fraction of galaxies in filaments closer to clusters\footnote{We remind the reader that our motivation for this study is observationally driven and therefore we chose to adopt a uniform thickness of the filaments. As stated in Sec. \ref{subsec:thickness}, we see in simulations that gas and dark matter filaments are getting thicker closer to nodes. Consequently, more halos should lie within filament boundaries closer to clusters. In our simplified convention tailored to observations, however, this additional factor will not be considered.}.
\begin{figure}
   \centering
   \includegraphics[width=\columnwidth]{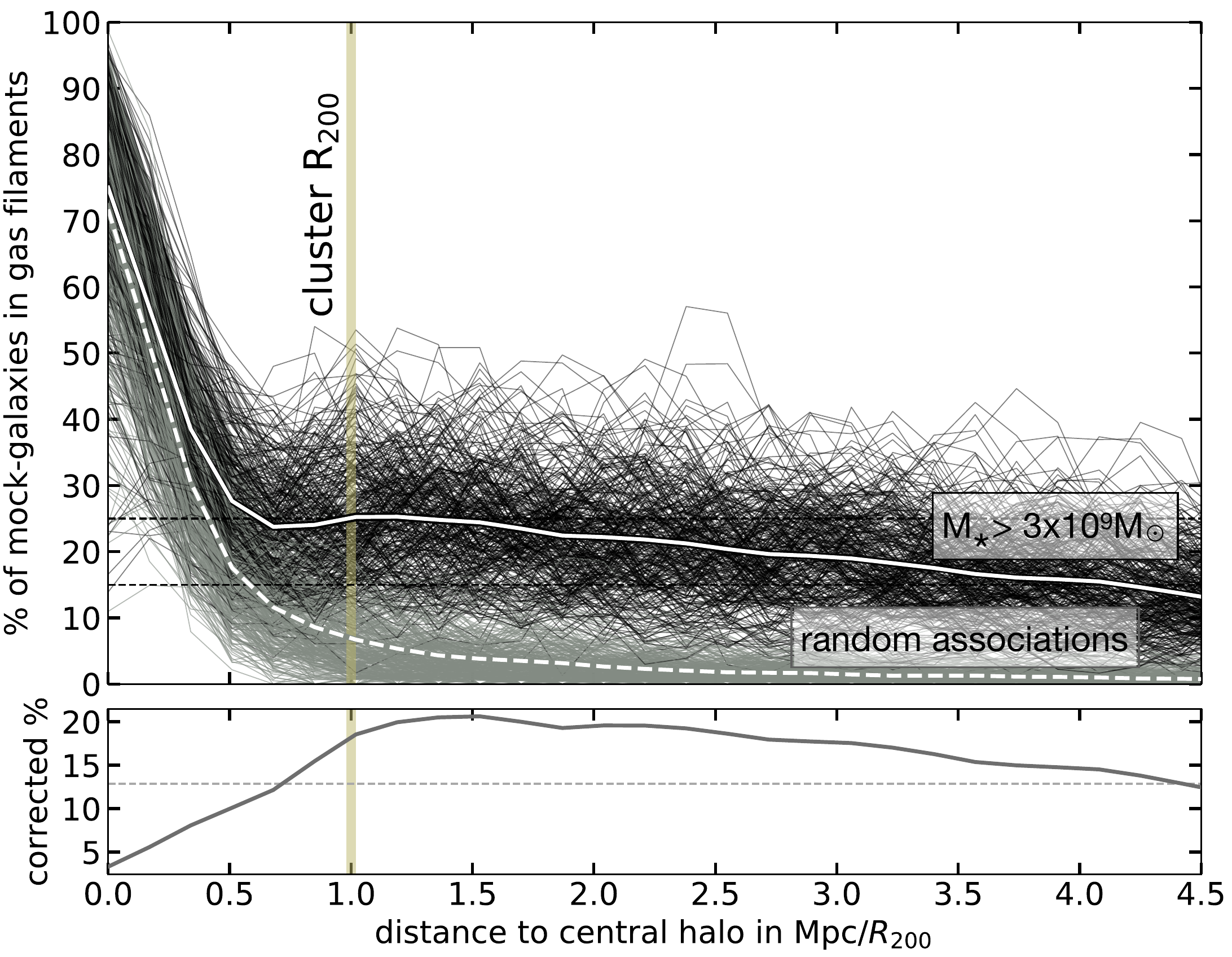}
    \caption{Percentage of mock galaxies in gas filaments ( $D_{\rm{skel}}<0.7 \hMpc$) as a function of radius for 324 clusters from \threehundred project at $z = 0$ (black lines and solid mean), normalised by $R_{200}$. Grey lines show the percentage of random associations to filaments. Lines converge inside $R_{200}$ where filaments are closer together than they are thick. The corrected percentage of galaxies in filaments is plotted in the lower panel. The percentage of galaxies in filaments increases from $\sim13$\% at the edge of the box to $\sim21$\% at $\sim1.5\,R_{200}$.}
   \label{fig:frac_dist2centre}
\end{figure}
In Figure \ref{fig:frac_dist2centre} we show the mean percentage of mock galaxies in gas filaments (with $D_{\rm{skel}}<0.7 \rm{Mpc}$) as a function of radius in steps of 500 pc (black lines). Going from the edge of the box to the cluster's $R_{200}$, we see that the fraction of mock galaxies belonging to filaments increases by about 10\% from $\sim 15\%$ to $\sim 25\%$. 
Closer to the cluster centre, the signal of the central halo is buried under the dominance of accumulating filaments in the small volume. Filaments are bunched together more closely than they are thick -- here, every galaxy will be near a filament. Therefore, inside $R_{200}$, the percentage of galaxies in filaments is rapidly approaching 100\%. At large scales, fractions resemble that of the cosmic average. 

The grey lines consider an important effect: even if the distribution of galaxies were random, some of them would still appear associated to filaments. This problem is particularly acute close to the cluster centres. We simulate this apparent association by randomising the angles of the filament networks. The dashed line shows the average percentage of galaxies within \mbox{$D_{\rm{skel}}<0.7\hMpc$} for these randomised filament networks.
This curve results from the combined effect of the growing number of galaxies and the increase in the fraction of the local volume occupied by filaments as we approach the cluster centres. By stacking all 324 clusters, this method allows us to correct for the random galaxy associations to filaments with high statistical accuracy.
\begin{figure*}
   \centering
   \includegraphics[width=\textwidth]{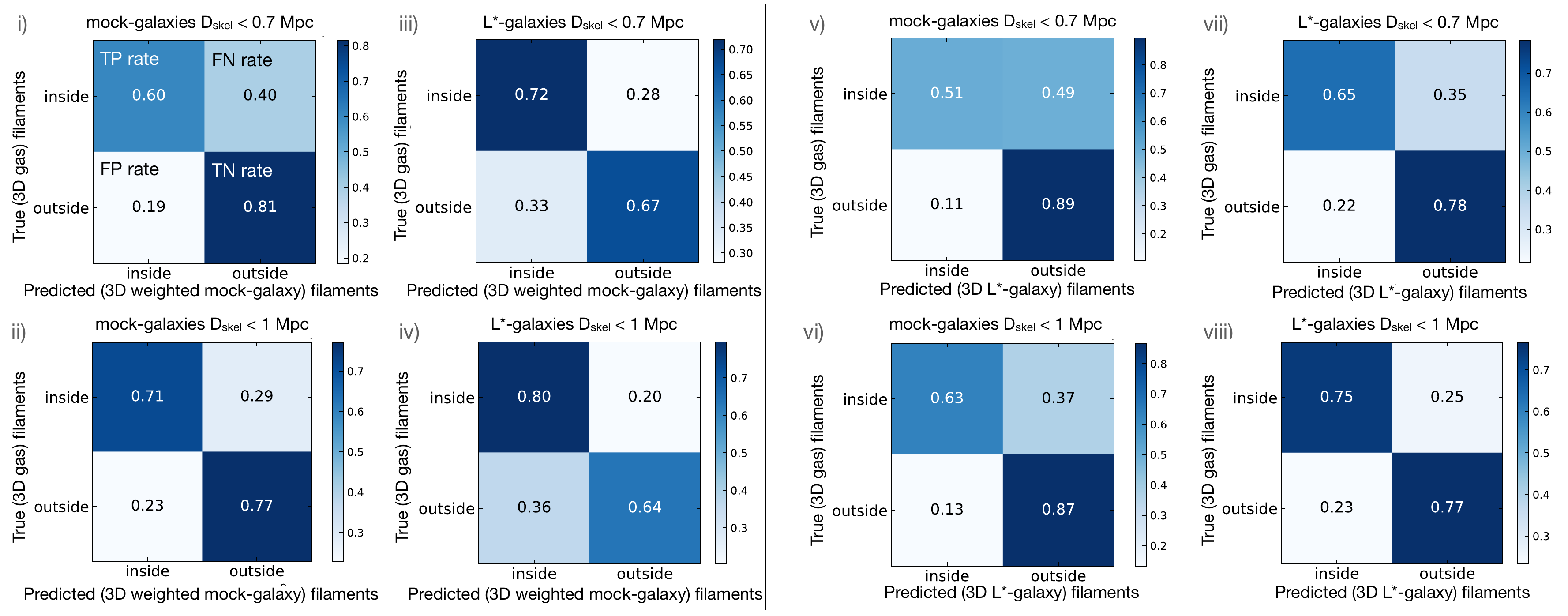}
    \caption{The confusion matrices (CM) document and evaluate the performance of associating galaxies to filaments in several 3D cases. The classification model labels whether a galaxy is inside or outside a filament, using 3D gas-filaments as the "true value". Each CM assesses a different choice: (1) filament thicknesses; $D_{\rm{skel}}<0.7\hMpc$ (top row) vs. $D_{\rm{skel}}<1\hMpc$ (bottom row) (2) mass-limits for galaxy-based filament extractions; $M_*> 3 \times 10^9 \Msun$ (left panel) vs. $L^*$ = $M_*> 10^{10} \Msun$ (right panel), and (3) mass-limits for galaxies associated to filaments; $M_*> 3 \times 10^9 \Msun$ (first and third column) vs. $L^*$ = $M_*> 10^{10} \Msun$ (second and fourth column). TP stands for true positive, FN for false negative, FP for false positive and TN for true negative rates, see text for details.}
   \label{fig:cm}
\end{figure*}
The lower panel of Fig. \ref{fig:frac_dist2centre} shows the corrected percentages of mock galaxies in gas filaments. 
Very close to the centre of the cluster, the numbers of galaxies in filaments are meaningless since we cannot distinguish between galaxies in filaments from random associations. However, this problem declines quickly, and by $1.5\times R_{200}$ the number of galaxies truly associated with filaments dominates the expected number of random associations by a factor of 10. Beyond $1.5\times R_{200}$, the probability for galaxies to be randomly associated to filaments becomes negligible. 
The fraction of galaxies in filaments steadily increases with proximity to the cluster from the edge of the simulated box until $\sim1.5\times R_{200}$. Between $4.5\times R_{200}$ and $\sim1.5\times R_{200}$, the fraction increases significantly from 12.8\% to 20.6\%.
At $\sim1.5\times R_{200}$, a plateau is reached and the curve turns over. The fraction of galaxies in filaments apparently declines beyond this point, but this close to the cluster centre the fraction becomes meaningless.  We see a similar increase of galaxies in mock galaxy filaments, albeit less prominent and at higher values (with an increase of corrected fractions from $\sim21$\% to $\sim25.8$\%).\footnote{Note that we can only speak in general terms here and give average numbers. Due to the oblate nature of clusters and the preference of filaments to align with the major axis of the cluster (Sec. \ref{subsec:alignment}), we expect some anisotropic variations to exist among the cluster sample.}

We conclude that the presence of a cluster influences the number of galaxies in filaments in its vicinity. 
We speculate that this could be, at least in part, a consequence of the high fraction of backsplash galaxies in the region between $1$ and $2 R_{200}$ of the cluster.  \citet{Haggar2020} have shown that between 30\% and 70\% (depending on cluster relaxedness) of all the galaxies in this region are members of the backsplash populations. These are galaxies that have passed through the centre of the cluster and are now located in the region between $R_{200}$ and  $2R_{200}$. These galaxies may not be isotropically distributed, retaining some memory of their accretion direction, and thus showing some preference to be located near filaments. The association of backsplash galaxies to filaments is potentially interesting, but it exceeds the scope of this study and will be examined in more detail in a future paper.  

The pile-up of filament galaxies as we approach the clusters seen in Fig. \ref{fig:frac_dist2centre} indicates that the accretion onto filaments accelerates closer to the cluster. This analysis therefore allows us to go beyond a model of a pure spherical collapse (radially defined "cluster core", "infall region" and "field" regimes) and to characterise the cluster "infall regime" using filaments and their 3D structure as additional environmental information. Whether a cluster galaxy has been accreted through  filaments or not may affect its properties and evolution, and depend on its exact accretion history. Being able to make this distinction is therefore important, and we will explore this question in the future.

\subsection{Performance evaluation for observations}
\label{subsec:performance}

Ideally, mock galaxies belong to both, the "truth table", (i.e., our reference frame, where galaxies are associated to the gas filament network) 
as well as the "predicted table" (i.e., they are associated to the filament network established using galaxies). Beyond this wish, the decisions for narrower (e.g., $D_{\rm{skel}}<0.7 \hMpc$) or thicker (e.g., $D_{\rm{skel}}<1 \hMpc$) filaments, and for filaments based on a deeper (e.g., $M_*> 3 \times 10^9 \Msun$) or brighter (e.g., $M_*> 10^{10} \Msun$) sample depends on the availability of (observational) data and the scientific question being addressed. In the following section, we assess purity, completeness, accuracy and precision of the method and samples we introduced in this paper. By monitoring different realistic simulated cases we aim to offer practical decision-making support for selection strategies in observations.

\subsubsection{The impact of filament-detection methods on recovery rates} 
\label{subsec:recoveryCM}

The confusion matrices (CM) in Fig. \ref{fig:cm} document and evaluate the performance of our classification based on the two criteria of being inside or outside a filament network. In this test, we are interested in a binary classifier: either a galaxy is part of a filament ("inside") or it is not ("outside"). 
We use  galaxies outside $R_{200}$ of the entire cluster sample for these predictions and treat fractions of galaxies in our reference filament network as our truth table: "True (3D gas) filaments". We test two filament extractions: "Predicted (3D weighted mock galaxy) filaments" (left panel, figures i--iv) and "Predicted (3D $L^*$-galaxy) filaments" (right panel, figures v--viii). We further show filament associations for galaxy samples of two mass limits, fractions of mock galaxies (figures i, ii and v, vi) and fractions of $L^*$-galaxies (figures iii, iv and vii, viii), as well as two filament thicknesses (top rows for $D_{\rm{skel}}<0.7\hMpc$ and bottom rows for $D_{\rm{skel}}<1\hMpc$). 
 
Figure \ref{fig:cm} can help make choices appropriate for the reader's science objective. 
First, decreasing the thickness of filaments leads to a purer sample. The false positive (FP) rate of galaxies in mock galaxy filaments (i.e., galaxies that are measured as being in filaments that really are not) decreases from 23\% in thicker filaments ($D_{\rm{skel}}<1\hMpc$ figure ii) to 19\% in narrower filaments ($D_{\rm{skel}}<0.7\hMpc$, figure i). However, choosing thicker filaments means that larger volumes get covered, which also leads to an increase in completeness: The true positive (TP) rate (i.e., galaxies that are measured as being in filaments that truly are) increases from 60\% in narrower filaments to 71\% in thicker filaments. 
For many applications a low false positive rate, e.g. below 20\% -- and thus an increase in purity --  will be the desired goal. Therefore, in the case where purity is most important, we advise narrower filaments of the order of $0.7\hMpc$. If, however, the scientific question benefits from a more complete sample, we advise to choose defining thicker filaments of the order of $1\hMpc$. Put another way, the accuracy will be higher in narrower filaments (Accuracy\footnote{Accuracy = (TN + TP)/total number} = 78\% vs. 75\%), but the precision\footnote{Precision = TP/(TP +FP)} will be lower (Precision = 43\% vs. 57\%). The method, however, stays the same. In this paper we have explicitly discussed the effect of the thickness on the results of our analysis whenever relevant. 

Figures iii) and iv) show results for $L^*$-galaxies in mock galaxy extracted filaments. Of all cases evaluated, this case reached the highest completeness rates: 72\% of all $L^*$-galaxies lie within narrow filaments (iii), increasing to 80\% for the thicker filament (iv) i.e., only 28\% (20\%) of $L^*$ galaxies are missed. 
However, the purity suffers. The false positive rate of 33\% for $D_{\rm{skel}}<0.7\hMpc$ and 36\% $D_{\rm{skel}}<1\hMpc$ is the highest of all cases. A quarter of all galaxies predicted to lie inside filaments are actually outside.
\begin{figure}
   \centering
   \includegraphics[width=\columnwidth]{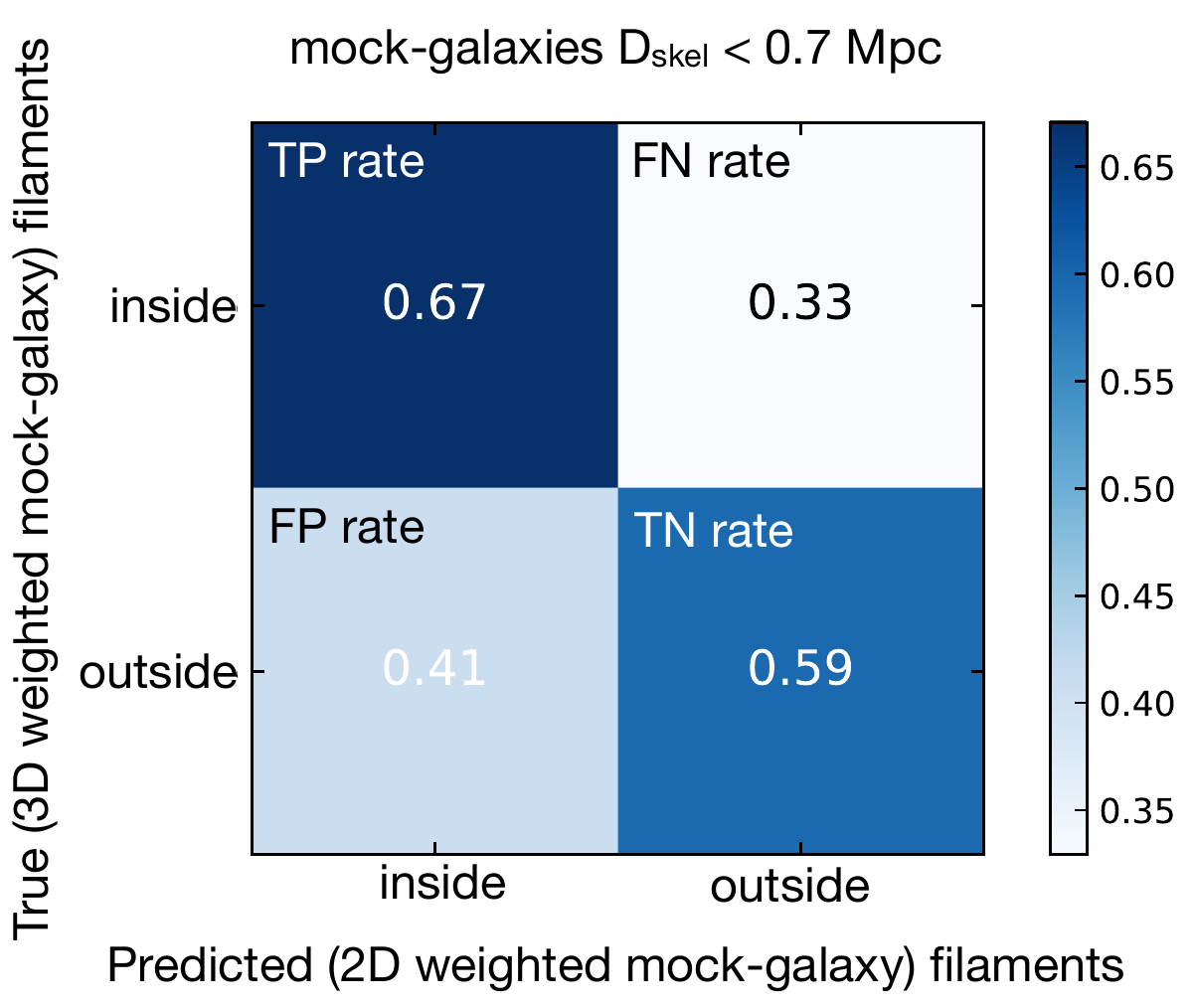}
    \caption{The confusion matrix describes the performance of associating mock galaxies to filaments in 2D projection. The classification model we use here labels whether a halo is inside or outside a filament ($D_{\rm{skel}}<0.7\hMpc$), using mock galaxies in 3D for the extraction as "true values". The high false positive rate (0.41) is largely the result of mis-classifying projected foreground and background galaxies as part of filaments.}
   \label{fig:cm_2D}
\end{figure}

In addition, we assess the following question: even if deep data for filament extraction is available, is a network extraction based on high mass galaxies the better choice? Given that massive galaxies trace filaments, this is a reasonable question to ask.
Taking this argument to an extreme case helps to underpin this issue: suppose only the most massive galaxies trace and shape filaments and low mass galaxies are uniformly distributed, then using the entire sample to find filaments is counterproductive. 
In an observational setup, low mass galaxies are also hardest to robustly classify as members of the structure and therefore they will have the highest membership contamination. 
The right-hand-side of figure \ref{fig:cm} evaluates this possibility. 
In the network that was extracted using $L^*$-galaxies, only half of all mock galaxies that actually are in filaments are recovered (v). This increases to 63\% for thicker filaments (vi). However, this offers relatively little contamination (FP rates of 11\%  and 13\%, accuracy of 82\% and 80\% for thinner and thicker filaments respectively).
Recovering $L^*$-galaxies in $L^*$-filaments yields both high accuracy (75\% and 76\%) and precision (58\% and 73\% for thinner and thicker filaments respectively).


In this paper, we have chosen to highlight the case of (weighted) mock galaxies with $M_*> 3 \times 10^9 \Msun$ -- for both filament extraction and halo association -- and of thinner filaments $D_{\rm{skel}}<0.7\hMpc$. Both of these choices are motivated by the WEAVE detection limit, our wish to study galaxies to lower mass limits and a preference for low false positive rates (less than 20\%, i.e., only 1 in 5 falsely classified as galaxies in filaments). In practice, this means the assessment helped to make choices appropriate for our science objective, for which we aim to maximise the contrast by choosing the purest sample.

\subsubsection{The impact of projections on recovery rates}
\label{subsec:impact}

Moving closer to realistic observational conditions, we test if galaxy rates associated to filaments extracted in three dimensions may be recovered in a two dimensional projection.
Our final question therefore is: what fraction of galaxies that are in filaments in 3D can we recover in 2D? 
Figure \ref{fig:cm_2D} shows the confusion matrix using mock galaxies around $0.7\,\rm{Mpc}$ of 3D mock galaxy filaments as the "truth table" and in 2D as "predicted  values". This corresponds to our favoured selection criterion introduced in Fig. \ref{fig:cm}i.

Because in 2D, the same number of 3D galaxies are projected onto a plane, there are apparently more galaxies close to filaments. We can therefore assume a high contamination rate for galaxies in filaments extracted from a two-dimensional projection of galaxies without any additional information. Fig. \ref{fig:cm_2D} shows that in 2D, we predict twice as many galaxies in filaments that actually are not than if we had 3D information (false positive rate of 0.41 in 2D vs. 0.19 in 3D for thinner filaments and 0.48 in 2D vs. 0.23 in 3D for thicker filaments).
That means that even in the case of well identified filaments, still half the galaxies are actually background or foreground galaxies. 
However, we still correctly identify 67\% (75\% for thicker filaments) of galaxies in filaments in 2D. So the true positive rate or completeness is still relatively high compared to if we randomly selected galaxies. A random selection of galaxies would only yield a true positive rate of 14\% (same as false positive rate) compared to 67\% if we select filaments in 2D. So while 2D filament extraction has its drawbacks in comparison to the full 3D information, it still improves the hit-rate by almost five times in comparison to a random selection. 

We remind the reader that these tests were performed in a controlled volume of a sphere with $15\hMpc$ radius around the cluster. The biggest remaining issue in an observational framework will be to remove foreground and background galaxies. One way of doing this is by identifying the volume of interest through spectroscopic redshifts. This will be the path for ensuring a clean sample for the upcoming WEAVE Wide-Field Cluster Survey where we expect between 4000--6000 spectroscopically identified cluster structure members out to $5R_{200}$.










\section{Conclusions}
\label{sec:conclusion}

Filaments are regarded as a crucial pathway for transporting matter into galaxy clusters.
While the cores and virialised regions of galaxy clusters and groups have been studied in detail, we must remember that the vast majority of galaxies spend significant time in large-scale  filaments and in infall regions that feed clusters. The outskirts of clusters are the regions where the infall and virtualisation of matter takes place, which is why future explorations are designed to map, characterise and study the large-scale structure in the outer envelopes of galaxy clusters \citep{Walker2019}. Understanding how galaxy properties are affected by the geography of their environmental history depends largely on how accurately and effectively we are able to map this geography. Due to the low density contrast outside $R_{200}$ in cluster regions, measurements are very challenging. It is therefore vital to test filament finding on simulated clusters that mimic the observations.  

We have used \threehundred project simulation suite to map and characterise filamentary structures around 324 massive simulated galaxy clusters. We extended our investigation from gas tracers to mock galaxies, and finished with an outlook for observational setups of future surveys, specifically highlighting the WEAVE Wide-Field Cluster Survey (WWFCS). We used realistic halo catalogues to quantify our ability to trace filaments from 2D observations limited to the immediate surroundings of clusters out to $5R_{200}$.

The main findings of this work are:

\textit{Simulations}
\begin{enumerate}[{(1)}]
\item We are able to reconstruct the filamentary distribution surrounding cluster out to $5R_{200}$, taking into account realistic observational limitations. Using the topological filament finder \disperse\, \citep{Sousbie2011} for the extraction, we establish the filamentary network around clusters based on smoothed gas particles as our reference framework.
\label{item:1}
\item Gas filaments align with the shape of the central (most massive) halo. Specifically, filaments preferentially align with the major axis of the cluster, and do so more prominently in elongated clusters. We also identify strong bridges between the halo and the second most massive halo.
\label{item:2}
\item Based on gas particle density profiles, we find that a constant filament thickness of $0.7\hMpc$ radius is a reasonable choice. However, changing this to a more relaxed $1\hMpc$ thickness -- as was used by some authors in the literature -- does not make a very large difference to our methods and results, and present and assess results for both values when relevant. \label{item:3}
\end{enumerate}

\textit{Towards observations}
\begin{enumerate}[{(1)}]
\setcounter{enumi}{3}
\item Using the filamentary network constructed from the gas particles as reference, we find that we are able to reliably extract filaments in 3D using mock galaxies based on simulated halos with $M*>3\times10^9 \Msun$, tailored to the mass-limit and expected numbers of the upcoming WWFCS. This is achieved by applying mass-weighting to the mock galaxy distribution as part of the extraction process. We are also able to reconstruct the filament network with reasonable accuracy using a higher mass limit of $M*>1\times10^{10} \Msun$, corresponding to the $\sim L^*$ limited samples already available in existing cluster surveys.   
\label{item:4}
\item We find that filament extraction from millions of simulated gas particles to thousands of simulated halos impacts the reliability of filament extraction more than the projection from a 3D halo distribution to projected 2D distribution: filaments are recovered well in 2D compared to 3D.
\label{item:5}
\item Filaments occupy only a small fraction (a few percent) of the entire simulated volume outside $R_{200}$, but a quarter of all mock galaxies with $M_*>3 \times 10^9 \Msun$ are in filaments (with a distance to filament ridges $D_{\rm{skel}}<0.7 \hMpc$). Normalised by the volume the filaments occupy, between 12\% and 14\% of mock galaxies lie in filament, depending on extraction method.    
\label{item:6}
\item The fraction of mock galaxies in filaments is independent of the mass or dynamical status of the central cluster, but depends on the mass-limit of the mock galaxy samples. For a given filament extraction method, more massive galaxies are more likely to be in filaments.
\label{item:7}
\item The presence of a cluster influences the number of galaxies in filaments in its vicinity. The fraction of galaxies in gas filaments increases from $\sim13$\% at $5R_{200}$ to a maximum of $\sim20.5$\% at $1.5R_{200}$. 
\label{item:8}
\item We present a set of confusion matrices that can help to choose appropriate selection criteria for filament extractions. If the goal is a maximally pure sample, it is better to define thinner filaments and extract filaments using a galaxy sample with a relatively low mass limit. This is harder to achieve closer to the cluster, where it is difficult to tell whether a galaxy is in or out of the converging filament network. If the scientific question benefits from a more complete sample, it is better to define thicker filaments. 
\label{item:9}
\item In observations, i.e., projected 2D space, we correctly identify 67\% (75\% for thicker filaments) of halos in filaments. In comparison, only 14\% of randomly selected galaxies lie in filaments. The methods presented here are therefore five times more efficient than a random selection of galaxies. 
\label{item:10}
\end{enumerate}

The approach presented in this paper allows to go beyond the traditional environmental regimes of \textit{cluster core}, \textit{infall region}, and \textit{field} -- which is based on a spherical collapse model. 
As we departure from sphericity, the cluster's region of influence is manifested by the facts that (1) the central halo itself is not spherical, (2) the accretion shock and backsplash galaxies are likely distributed in preferential directions \citep{Haggar2020}, (3) the cosmic filaments are connected to the cluster in preferential directions (section \ref{subsec:alignment}) and (4) galaxies preferentially lie in filaments (section \ref{subsec:association}). Combined, this leads to an increasingly non-spherical appearance of the cluster. 
In addition, the tracers that form filaments are biased in the sense that more massive galaxies lie preferentially in the vicinity of filaments. 

Applied to future observations, our method provides the groundwork for successful realisations of research projects that involve the analysis and interpretation of a new generation of galaxy evolution experiments.



\section*{Acknowledgements}
\textit{This work has been made possible by \threehundred collaboration\footnote{https://www.the300-project.org}, which benefits from financial support of the European Union’s Horizon 2020 Research and Innovation programme under the Marie Sk\l{}odowskaw-Curie grant agreement number 734374, i.e. the LACEGAL project. \threehundred simulations used in this paper have been performed in the MareNostrum Supercomputer at the Barcelona Supercomputing Center, thanks to CPU time granted by the Red Espa\~{n}ola de Supercomputaci\'{o}n.
UK acknowledges support from STFC, 
WC acknowledges supported from the European Research Council under grant number 670193. CM acknowledges support from the China Scholarship Council (CSC). AK is supported by MICIU/FEDER under research grant PGC2018-094975-C21. He further acknowledges support from the Spanish Red Consolider MultiDark FPA2017-90566-REDC and thanks Ride for nowhere.
\newline
The authors contributed to this paper in the following ways: UK, AAS, MEG, and FRP formed the core team. UK ran \disperse, analysed the data, produced the plots and wrote the paper, with the following exceptions: CW ran \disperse\ on the gas particles; Sec. \ref{subsec:thickness} is a summary of work done by AR, Sec. \ref{subsec:alignment} is a summary of work done by CM; RH calculated the relaxedness parameter $R$ based on dynamical state parameters produced by WC; GY supplied the simulation data; AK the halo catalogs.}




\bibliographystyle{mnras}
\bibliography{references.bib} 

\
\bsp	
\label{lastpage}
\end{document}